\documentstyle[12pt,psfig,axodraw]{article}

\begin{document}
\baselineskip 0.7cm

\newcommand{\gsim}{ \mathop{}_{\textstyle \sim}^{\textstyle >} }
\newcommand{\lsim}{ \mathop{}_{\textstyle \sim}^{\textstyle <} }
\newcommand{\EV}{ {\rm eV} } 
\newcommand{\KEV}{ {\rm keV} }
\newcommand{\MEV}{ {\rm MeV} } 
\newcommand{\GEV}{ {\rm GeV} }
\newcommand{\TEV}{ {\rm TeV} }
\renewcommand{\thefootnote}{\fnsymbol{footnote}}
\setcounter{footnote}{1}

\begin{titlepage}
\today
\begin{flushright}
UT-784
%\\
\end{flushright}

\vskip 0.35cm
\begin{center}
  {\large \bf Phenomenological Aspects of a Direct-transmission Model of
    Dynamical Supersymmetry Breaking with the Gravitino Mass 
    $m_{3/2} < 1~\KEV$} 
  \vskip 1.2cm Yasunori~Nomura
  and Kazuhiro~Tobe,\footnote{Fellow of the Japan Society for the
    Promotion of Science.}\footnote{Address 
    after September 1 : Department of Physics,
    The Ohio State University, Columbus, Ohio 43210}
 \vskip 0.4cm

  {\it Department of Physics, University of Tokyo,\\ Bunkyo-ku,
    Hongou, Tokyo 113, Japan}

  \vskip 1.5cm
  
  \abstract{We analyze a direct-transmission model of dynamical
    SUSY breaking previously proposed. In the model
    the gravitino mass is naturally smaller than $1~\KEV$, which is
    required from the standard cosmology. We find that there are many
    distinguishable features from in other models: 
    for example the so-called GUT
    relation among the gaugino masses does not hold even if we
    consider the GUT models. Furthermore, the gauginos are always
    lighter than the sfermions since the gaugino masses have extra
    suppression factors.
    We also discuss a collider signature ``$\gamma \gamma +$ 
    missing energy'' in the present model.}
\end{center}
\end{titlepage}

\renewcommand{\thefootnote}{\arabic{footnote}}
\setcounter{footnote}{0}

%
%
%       *** Main Part ***
%
%

\section{Introduction}

Low-energy supersymmetry (SUSY) is very attractive since it 
stabilizes a hierarchy between the weak scale and a higher scale
of new physics (say the Planck scale). 
It is also strongly supported by the observed 
unification of the standard model (SM) gauge couplings.

Recently, low-energy dynamical SUSY breaking (DSB) with gauge mediation has
attracted attention since it may provide a natural
explanation of large hierarchy between the weak scale and the
Planck scale as well as a natural solution to the SUSY
flavor changing neutral current (FCNC) and the CP problems
\cite{gm,Dine_review}.

Several mechanisms for DSB have been discovered
\cite{ADS,ISS,IY,Randall} and their applications to realistic
models have been proposed in the literatures~\cite{DNS,HIY,MH}.  
Most of the
models which have been proposed, however, need 
relatively large DSB scales $\Lambda~\gsim~10^7~\GEV$ to get sufficiently
large SUSY breaking masses in the minimal supersymmetric 
standard model (MSSM) sector. As a result, the gravitino
mass becomes $m_{3/2} > 10~\KEV$.\footnote{In 
the model in Ref. \cite{HIY} it seems possible
 to make the gravitino mass smaller than $1$ keV due to strong gauge
 interaction.  If a naive dimensional analysis of strongly coupled SUSY
 theory \cite{NT} is applicable, however, the gravitino mass is 
 likely larger than $1$ keV \cite{NTY}.}
On the other hand, 
the gravitino should be lighter than $1~\KEV$
so that it does not overclose the universe
\cite{gravitino_mass}. To escape this bound, 
one may consider a late-time entropy production 
which, however, leads to a complicated cosmology \cite{MM,MMY}.  
Furthermore, as
recently pointed out in Ref. \cite{HKY}, a constraint on the
cosmic X($\gamma$)-ray backgrounds from 
a dilaton decay requires that the gravitino should
be lighter than $100~\KEV$. To get rid of this bound, we have to
construct a string theory without the dilaton.  Such a string theory,
however, has not been known. Therefore, it is very important to
construct a model with $m_{3/2} < 1~\KEV$ and to investigate
phenomenological consequences of such a model.

In Ref. \cite{INTY}, a DSB model which transmits SUSY breaking effects
to the MSSM sector directly has been proposed. In this model, we can
lower the SUSY breaking scale and hence the gravitino mass can be set 
$m_{3/2} < 1~\KEV$ to avoid an introduction of complicated
non-standard cosmology keeping the advantage of gauge
mediation.  Moreover, such a light gravitino
may suggest a distinct signature in the existing collider experiments, that
is, the next-to-lightest superparticle (NLSP), mostly bino, can decay
into the gravitino, which is the lightest superparticle (LSP), within
detectors producing a observable ``$\gamma \gamma +$ missing energy'' signal 
as discussed in many literatures \cite{Rad_sym_break_MMM1,
gg_missing}.  Even if the signature  `` $\gamma \gamma +$ missing energy''
is not observed within the detectors, the slow decay of the NLSP may be 
detectable in the near future experiments such as LHC as pointed out 
in Ref. \cite{Orito}.
Furthermore, in the model in Ref. \cite{INTY}, the mass
spectrum of superparticles in the MSSM sector is quite different
from in the ordinary gauge mediation models and in the gravity mediation
models based on supergravity. In particular, the grand unified theory (GUT) 
relation among the gaugino
masses does not hold even if we consider the GUT unification.
Since the present model has many different features from the 
other ordinary models \cite{DNS,MH,HIY}, it
may be distinguishable.

The purpose of this paper is to investigate the mass spectrum in the
model in Ref. \cite{INTY} imposing experimental constraints and to
show the existence of the phenomenologically viable parameter 
regions in which the
gravitino mass is smaller than $1$ keV.  This paper is organized as
follows.  In Section 2, we briefly review the model in
Ref. \cite{INTY}.  In Section 3, we consider the low-energy mass
spectrum of the gauginos and sfermions in the MSSM sector, and argue
their typical features. The radiative electroweak symmetry breaking 
is also discussed.
In Section 4, we analyze the GUT models with and without the Yukawa
unification.  In section 5, we discuss other interesting features in
the present model. Section 6 is devoted to our conclusions.

\section{A direct-transmission model of SUSY breaking}

In this section, we review a model which has been proposed in
Ref. \cite{INTY}.  Let us first consider a dynamics for scale
generation.  To generate the scale, we adopt a SUSY SU(2) gauge theory 
with four doublet chiral superfields $Q_i$, where $i$ is a flavor index
($i=1,\cdots,4$).  Without a superpotential, this theory has a flavor
SU(4)$_F$ symmetry.  This SU(4)$_F$ symmetry is explicitly broken down
to a global SP(4)$_F$ by a superpotential in this model.  We add gauge
singlets $Y^a$ ($a=1, \cdots, 5$) which constitute a five-dimensional
representation of SP(4)$_F$
to obtain a tree-level superpotential
\begin{eqnarray}
 W_Y = \lambda_Y Y^a (QQ)_a,
\label{tree_superpotential}
\end{eqnarray}
where $(QQ)_a$ denote a five-dimensional representation of SP(4)$_F$
given by a suitable combination of gauge invariants $Q_iQ_j$.

A low energy effective superpotential with 
$W_Y$ in Eq.(\ref{tree_superpotential})
, which describes the dynamics of the SU(2) gauge interaction, 
may be given by \cite{IS}
\begin{eqnarray}
 W_{eff}=S(V^2 + V_a^2 - \Lambda^4) + \lambda_Y Y^a V_a
\label{dynamical_potential}
\end{eqnarray}
in terms of low-energy degrees of freedom
\begin{eqnarray}
 V \sim (QQ), \quad V_a \sim (QQ)_a,
\end{eqnarray}
where $S$ is an additional chiral superfield, $\Lambda$ a
dynamically generated scale, and a gauge invariant ($QQ$) denotes a
singlet of
SP(4)$_F$ defined by
\begin{eqnarray}
(QQ)=\frac{1}{2} (Q_1 Q_2 + Q_3 Q_4).
\end{eqnarray}
The effective superpotential Eq.(\ref{dynamical_potential}) implies
that the singlet $V \sim (QQ)$
condenses as 
\begin{eqnarray}
\label{VEV}
 \langle V \rangle = \Lambda^2,
\end{eqnarray}
and SUSY is kept unbroken in this unique vacuum.  Since the vacuum
preserves the flavor SP(4)$_F$ symmetry, we have no massless
Nambu-Goldstone boson. The absence of flat direction at this stage is
crucial for causing dynamical SUSY breaking as seen below.

Next we consider the dynamical SUSY breaking \cite{IY}.  
Let us first consider
a model with one singlet chiral superfield $Z$ for SUSY breaking
which couples directly to $(QQ)$.
Then, the superpotential is given by
\begin{eqnarray}
  W=W_Y+ \lambda Z (QQ).
\end{eqnarray}
For a relatively large value of the coupling $\lambda_Y$, we again
obtain the condensation Eq.(\ref{VEV}) with low-energy
effective superpotential approximately given by
\begin{eqnarray}
  W_{eff} \simeq \lambda \Lambda^2 Z.
\end{eqnarray}
Then, $F_Z \simeq \lambda \Lambda^2 \neq 0$ and SUSY is broken.

On the other hand,
the effective K\"ahler potential is expected to take a form
\begin{eqnarray}
  K = |Z|^2 -
     \frac{\eta}{4 \Lambda^2}|\lambda Z|^4 + \cdots,
\end{eqnarray}
where we expect that $\eta$ is a real constant of order one.  Then the
effective potential for the scalar $Z$ (with the same notation as
the superfield) is given by
\begin{eqnarray}
  V_{eff} \simeq |\lambda|^2 \Lambda^4 (1 +
         \frac{\eta}{\Lambda^2} |\lambda|^4 |Z|^2).
\end{eqnarray}
If $\eta > 0$, this implies $\langle Z \rangle = 0$.  Otherwise we
expect $|\lambda \langle Z \rangle| \sim \Lambda$, since the effective
potential is lifted in the large $|Z|$ ($> \Lambda$) region
\cite{IY,HIY,Shirman}.  

In the following analyses, we assume the
latter case $|\lambda \langle Z \rangle| \sim \Lambda$, which results
in the breakdown of the $R$ symmetry.\footnote{
Even if we take $|\lambda \langle Z \rangle| \sim 4\pi\Lambda$ 
the main conclusion in the present paper does not change.}
  The spontaneous breakdown of the
$R$ symmetry produces a Nambu-Goldstone $R$-axion. This $R$-axion is,
however, cosmologically harmless, since it acquires a mass from the
$R$-breaking constant term in the superpotential which is necessary to
set the cosmological constant to zero \cite{Bag}.\footnote{
When $\langle Z \rangle = 0$, 
appropriate $R$-breaking mass terms such as $md{\bar d}
+ m'l{\bar l}$ is necessary to give masses to the MSSM 
gauginos because the $R$ symmetry keeps the gauginos massless.} 

To transmit the SUSY breaking effects to the MSSM sector, we introduce
vector-like messenger quark multiplets $d$, $\bar{d}$ and lepton
multiplets $l$, $\bar{l}$. We assume that the $d$ and $\bar{d}$
transform as the right-handed down quark and its antiparticle,
respectively, under the SM gauge group.  The $l$ and $\bar{l}$ are
assumed to transform as the left-handed lepton doublet and its
antiparticle, respectively.\footnote{One may consider that the messenger
  quark and lepton multiplets are embedded into SU(5) GUT multiplets
  $\bf 5$ and $ {\bf 5}^*$ 
  to preserve the unification of the SM gauge coupling constants.} We
introduce coupling of the messenger quarks and leptons to the singlet
$Z$ in order to directly transfer the SUSY breaking to the messenger sector.
Then the effective superpotential is 
\begin{eqnarray}
  W_{eff} \simeq Z ( \lambda \Lambda^2 +
  k_d d \bar{d} + k_l l \bar{l}).
\end{eqnarray}
In this case, however, the condensations $\langle d \bar{d} \rangle
\neq 0$ and $\langle l \bar{l} \rangle \neq 0$ occur, and hence SUSY
is not broken ($F_Z=0$).  To avoid such undesired condensations, 
we further introduce a pair of messenger quark and lepton
multiplets $(d',l')+(\bar{d}',\bar{l}')$, and mass parameters
$m_d$, $m_{\bar{d}}$, $m_l$, and $m_{\bar{l}}$ as follows:
\begin{eqnarray}
  W_{eff}&=&Z (\lambda \Lambda^2 + k_d d \bar{d} + k_l l \bar{l})
  \nonumber \\ &&+ m_d d \bar{d}' + m_{\bar{d}} d' \bar{d}
+m_l l \bar{l}' + m_{\bar{l}} l' \bar{l}.
\label{one_singlet_superpotential}
\end{eqnarray}
Dynamical generation of these mass parameters $m_d$, $m_{\bar{d}}$,
$m_l$, and $m_{\bar{l}}$ has been discussed in
Ref. {\cite{INTY}}. We will briefly review it below.
  
Owing to the mass parameters in Eq.(\ref{one_singlet_superpotential}),
we can obtain a SUSY breaking vacuum with vacuum expectation
values of the messenger squarks and sleptons vanishing,
\begin{eqnarray}
  \langle d \rangle = \langle \bar{d} \rangle=\langle l \rangle
  =\langle \bar{l} \rangle=\langle d' \rangle=\langle \bar{d}' \rangle
  =\langle l' \rangle=\langle \bar{l}'\rangle=0, \quad
\langle F_Z \rangle \simeq \lambda \Lambda^2.
\end{eqnarray}
The condition for this desired vacuum to be the true one
is given by examining the scalar potential,
\begin{eqnarray}
  V&=&| \lambda \Lambda^2 + k_d d \bar{d}+ k_l l \bar{l} |^2
%\nonumber \\
%&&
  + |m_{\bar{d}}\bar{d}|^2 + |m_d d|^2 + |m_{\bar{l}}\bar{l}|^2 + |m_l
  l|^2 \nonumber \\ && + |k_d Z \bar{d} + m_d \bar{d}'|^2
+|k_d Z d + m_{\bar{d}} d'|^2
%\nonumber \\
%&&
+ |k_l Z \bar{l} + m_l \bar{l}'|^2
+|k_l Z l + m_{\bar{l}} l'|^2,
\end{eqnarray}
as follows:
\begin{eqnarray}
  |m_d m_{\bar{d}}|^2 &>& |k_d \langle F_Z \rangle|^2 , \nonumber \\ 
|m_l m_{\bar{l}}|^2 &>& |k_l \langle F_Z \rangle|^2.
\label{stable_cond}
\end{eqnarray}
Then, the soft SUSY breaking masses of the messenger squarks and
sleptons are directly generated by $\langle F_Z \rangle \simeq \lambda
\Lambda^2$ through the couplings $Z(k_d d \bar{d} + k_l l \bar{l})$.
We will see later that such a direct-transmission 
of the SUSY breaking effects to
the messenger sector makes it possible to realize the light gravitino
$m_{3/2} < 1~\KEV$, which is required from the standard cosmology.

Now we discuss the dynamical generation of the mass parameters $m_d$,
$m_{\bar{d}}$, $m_l$, and $m_{\bar{l}}$.  To generate these mass
parameters dynamically, we introduce a singlet $X$ whose vacuum
expectation value plays the role of mass parameters.  Furthermore,
in order to give the vacuum expectation value to $X$ keeping the SUSY 
breaking, three singlet chiral supermultiplets $Z_i$ $(i=1,..,3)$ which
couple to $(QQ)$ are introduced as follows:\footnote{ We can construct a
  model in which only two singlet chiral supermultiplets are needed to
  give the vacuum expectation value to $X$ keeping the SUSY breaking if
  the GUT unification of the Yukawa couplings holds at the GUT scale.
  However, we consider three singlets because the
  Yukawa coupling unification
is easily broken by non-renormalizable operators as discussed later.}
\begin{eqnarray}
  W &=& W_Y + Z_1(\lambda_1(QQ) + k_{d1}d{\bar d} + k_{l1}l{\bar l} -
  f_1X^2) + Z_2(\lambda_2(QQ) + k_{d2}d{\bar d} + k_{l2}l{\bar l} -
  f_2X^2) \nonumber \\ &&+ Z_3(\lambda_3(QQ) + k_{d3}d{\bar d} +
  k_{l3}l{\bar l} - f_3X^2) + X(f_d d {\bar d}' + f_{\bar d} d' {\bar
    d} + f_l l {\bar l}'
     + f_{\bar l} l' {\bar l}).
\label{superpotential_model3}
\end{eqnarray}
Here, we should stress that the superpotential
Eq.(\ref{superpotential_model3}) is natural, since it has a global
symmetry U(1)$_R \times$U(1)$_\chi$, where U(1)$_R$ is an $R$
symmetry. That is, the superpotential Eq.(\ref{superpotential_model3})
is a general one allowed by the global U(1)$_R
\times$U(1)$_\chi$.\footnote{ This global symmetry may forbid mixings
  between the messenger quarks and the down-type quarks in the
  SM sector. This avoids naturally the FCNC problem \cite{DNS2}. 
  Then there exists the lightest
  stable particle in the messenger sector \cite{DGP2}.} The charges for
chiral superfields
are given in Table~\ref{table_charge}.
\begin{table}
\begin{center}
\begin{tabular}{|c|c|c|}  \hline 
  & & \\ & $Q, \psi, \bar{\psi}, X$ & $Z_i, \psi', \bar{\psi}'$\
\\ \hline  
U(1)$_R$ & $0$ & $2$ \\ \hline U(1)$_\chi$ & $1$ & $-2$ \\ \hline
\end{tabular}
\end{center}
\caption{U(1)$_R\times$ U(1)$_\chi$ charges for chiral superfields.
  Here, $\psi=d,l$ and $i=1,2,3$.}
\label{table_charge}
\end{table}

Without loss of generality, we may set $k_{d1} = k_{l1} = f_2 = 0$ by
an appropriate redefinition of $Z_1$, $Z_2$, and $Z_3$.
Under the following condition:
\begin{eqnarray}
  |f_\psi f_{\bar{\psi}} (\lambda_1 f_1 + \lambda_3 f_3)|^{2} >
  |k_{\psi 2} \lambda_2 (f_1^{2}+f_3^{2}) + k_{\psi 3} f_1 (\lambda_3
  f_1 -
\lambda_1 f_3)|^{2}
\label{vacuum_stability_condition}
\end{eqnarray}
for $\psi=d,l,$ the superpotential yields a vacuum with
\begin{eqnarray}
  \langle X \rangle = \sqrt{\frac{\lambda_1 f_1 + \lambda_3 f_3}
{f_1^{2}+f_3^{2}}} \Lambda,
\label{vev_X}
\end{eqnarray}
and vacuum expectation values of the messenger squarks and sleptons
vanish. The condition Eq.(\ref{vacuum_stability_condition})
corresponds to the vacuum stability condition Eq.(\ref{stable_cond}).
In this vacuum, the $F$-components of $Z_i$ are given by
\begin{eqnarray}
  F_{Z_1} = \frac{\lambda_1 f_3 - \lambda_3 f_1}{f_1^{2}+f_3^{2}} f_3
  \Lambda^2, \quad F_{Z_2} = \lambda_2 \Lambda^2, \quad F_{Z_3} =
  \frac{\lambda_3 f_1 - \lambda_1 f_3}{f_1^{2}+f_3^{2}}
f_1 \Lambda^2,
\label{vev_fz}
\end{eqnarray}
and thus SUSY is broken.

Since the scalar component of $X$ superfield has the vacuum
expectation value $\langle X \rangle$, the mass parameters $m_d$,
$m_{\bar{d}}$, $m_l$,
and $m_{\bar{l}}$ are dynamically generated as
\begin{eqnarray}
  m_{\psi}&=&f_{\psi} \langle X \rangle,
\\
  m_{\bar{\psi}} &=& f_{\bar{\psi}} \langle X \rangle
\end{eqnarray}
for $\psi=d,l$. Therefore, this model is 
reduced to the model described in Eq.(\ref{one_singlet_superpotential})
effectively. In a practical analysis we use the reduced model 
with Eq.(\ref{one_singlet_superpotential}).
We should note that all of the mass parameters are
at the same order of the SUSY breaking scale $\sqrt{F_{Z_i}}$ if the
Yukawa couplings $f_i$, $\lambda_i$, $f_\psi$, $f_{\bar{\psi}}$ 
are $O(1)$ because they
are generated by the same dynamics with the scale
$\Lambda$.

The messenger sfermions receive the SUSY breaking mass squared as
$k_{\psi 2} \langle F_{Z_2} \rangle +k_{\psi 3} \langle F_{Z_3} \rangle$.
Therefore, the gauginos and sfermions in the MSSM sector acquire 
their masses through loop diagrams of the messenger 
multiplets \cite{DNS,Martin,DG}.
We will discuss the obtained
mass spectrum in the MSSM
sector in the next section.

\section{Mass Spectrum of the superparticles in the MSSM sector}
\subsection{Mass Spectrum}
\label{Mass_spectrum}

In this section, we derive the low-energy mass spectrum of the
gauginos, squarks, and sleptons in the MSSM sector.
To calculate the masses for the gauginos and sfermions, we first consider
the mass eigenstates of the messenger fermions and sfermions.  To
begin with, the superpotential for the mass terms of the
messenger fields $\psi$, $\bar{\psi}$, $\psi'$, and $\bar{\psi}'$ for
$\psi=d,l$ is
represented as
\begin{eqnarray}
W=\sum_{\psi=d,l}(\bar{\psi}, \bar{\psi}')
M^{(\psi)}
\left(
\begin{array}{c}
  \psi \\ \psi'
\end{array}
\right),
\end{eqnarray}
where the mass matrix $M^{(\psi)}$ is given by
\begin{eqnarray}
  M^{(\psi)}= \left(
\begin{array}{cc}
  m^{(\psi)} & m_{\bar{\psi}}\\ 
m_{\psi} & 0
\end{array}
\right).
\label{mass_matrix}
\end{eqnarray}
In the present model, the above mass parameters
are given by 
\begin{eqnarray}
  m^{(\psi)} &=& k_{\psi 2} \langle Z_2 \rangle + k_{\psi 3} \langle
  Z_3 \rangle,
\label{mass_to_Z}
\\
m_{\psi} &=& f_{\psi} \langle X \rangle ,
\\
m_{\bar{\psi}} &=& f_{\bar{\psi}} \langle X \rangle.
\end{eqnarray}
Then, the messenger quark and lepton masses are given by diagonalizing
the
mass matrix $M^{(\psi)}$ as follows:
\begin{eqnarray}
  {\rm diag} \left( M^{(\psi)}_1, M^{(\psi)}_2 \right) = U^{(\psi)}
  M^{(\psi)} V^{(\psi)\dagger}.
\end{eqnarray}
On the other hand, the messenger squarks and sleptons receive the soft
SUSY breaking mass terms,
\begin{eqnarray}
{\cal{L}}_{soft}&=&\sum_{\psi=d,l} F^{(\psi)}\tilde{\psi} \tilde{\bar{\psi}},
\end{eqnarray}
where
\begin{eqnarray}
  F^{(\psi)}=k_{\psi 2} \langle F_{Z_2} \rangle
+ k_{\psi 3} \langle F_{Z_3} \rangle.
\end{eqnarray}
Then, the mass terms of the messenger squarks and sleptons are written as
\begin{eqnarray}
  {\cal{L}}_{s}=-\sum_{\psi=d,l}(\tilde{\psi}^{*}, \tilde{\psi}'^{*},
\tilde{\bar{\psi}}, \tilde{\bar{\psi}}')
\tilde{M}^{2(\psi)}
\left(
\begin{array}{c}
  \tilde{\psi} \\ \tilde{\psi}' \\ \tilde{\bar{\psi}}^{*} \\ 
  \tilde{\bar{\psi}}'^{*}
\end{array}
\right),
\end{eqnarray}
where the mass matrix $\tilde{M}^{2(\psi)}$ is given by
\begin{eqnarray}
  \tilde{M}^{2(\psi)}= \left(
\begin{array}{cccc}
  |m^{(\psi)}|^{2}+|m_{\psi}|^{2} & m^{(\psi)*} m_{\bar{\psi}} &
  F^{(\psi)*} & 0 \\ m^{(\psi)} m_{\bar{\psi}}^* &
  |m_{\bar{\psi}}|^{2} & 0 & 0 \\ F^{(\psi)} & 0 &
  |m^{(\psi)}|^{2}+|m_{\bar{\psi}}|^{2} & m^{(\psi)} m_{\psi}^* \\ 
0 & 0 & m^{(\psi)*} m_{\psi} & |m_{\psi}|^{2}
\end{array}
\right)
\end{eqnarray}
for $\psi=d,l$.
This can be diagonalized by the unitary matrix $T^{(\psi)}$ as
\begin{eqnarray}
  {\rm diag} \left(
m^{{(\psi)}2}_1, m^{{(\psi)}2}_2,m^{{(\psi)}2}_3, m^{{(\psi)}2}_4 
\right)=
T^{(\psi)} \tilde{M}^{2(\psi)} T^{(\psi)\dagger}.
\end{eqnarray}
When we take the mass eigenstates of the messenger sector, the
interactions for the messenger fields are described in terms of the
mixing matrices $U^{(\psi)}$, $V^{(\psi)}$, and $T^{(\psi)}$. (See
Appendix A.) Then we calculate the masses for the superparticles in
the MSSM sector.

The MSSM gauginos acquire 
masses through the one-loop diagrams 
of the messenger quark and lepton multiplets 
shown in Fig.\ref{gaugino_mass_graph}.
Their masses are given by
\begin{eqnarray}
  m_{\tilde{g}_3} &=& \frac{\alpha_3}{2\pi} {\cal F}^{(d)},
\nonumber \\
m_{\tilde{g}_2} &=& \frac{\alpha_2}{2\pi} {\cal F}^{(l)},
\nonumber \\
m_{\tilde{g}_1} &=& \frac{\alpha_1}{2\pi} \left\{
\frac{2}{5} {\cal F}^{(d)} + \frac{3}{5} {\cal F}^{(l)} \right\},
\label{gaugino_mass}
\end{eqnarray}
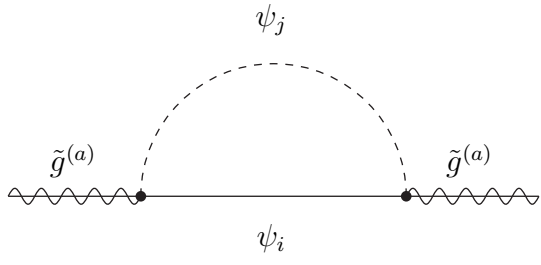
\begin{figure}
\begin{center} 
\begin{picture}(100,80)(150,120)
  \Line(100,150)(150,150) \Text(125,160)[b]{$\tilde{g}^{(a)}$}
  \Photon(100,150)(150,150){3}{5} \Line(150,150)(250,150)
  \Text(200,140)[t]{$\psi_i$} \DashCArc(200,150)(50,0,180){3}
  \Line(300,150)(250,150) \Text(275,160)[b]{$\tilde{g}^{(a)}$}
  \Photon(300,150)(250,150){3}{5} \Vertex(150,150){2}
  \Vertex(250,150){2}
\Text(200,215)[b]{${\tilde \psi}_j$}
%
%\put(175,50){\Huge{Fig.1}}
\end{picture}
\caption{Diagram contributing to the gaugino masses} 
\label{gaugino_mass_graph}
\end{center}
\end{figure}
where we have adopted the SU(5) GUT normalization of the U(1)$_Y$
gauge coupling ($\alpha_1 \equiv \frac{5}{3} \alpha_Y$), and
$m_{\tilde{g}_i}$ ($i=1, \cdots, 3$) denote the bino, wino, and gluino
masses, respectively. Here ${\cal F}^{(d)}$ represents
contributions from the messenger quark multiplets $d$, $d'$,
$\bar{d}$, and $\bar{d}'$, and ${\cal{F}}^{(l)}$ contributions
from the messenger lepton multiplets $l$, $l'$, $\bar{l}$, and
$\bar{l}'$.
The functions ${\cal F}^{(\psi)}$ $(\psi=d,l)$ are given by 
\begin{eqnarray}
  {\cal F}^{(\psi)} &=& \sum_{\alpha = 1}^{2} \sum_{\beta = 1}^{4}
  M_{\alpha}^{(\psi)} (U_{\alpha 1}^{(\psi)}T_{3 \beta}^{(\psi)
    \dagger} + U_{\alpha 2}^{(\psi)}T_{4 \beta}^{(\psi) \dagger})
  (T_{\beta 1}^{(\psi)}V_{1 \alpha}^{(\psi) \dagger} + T_{\beta
    2}^{(\psi)}V_{2 \alpha}^{(\psi) \dagger}) \nonumber \\ & & \times
  \frac{m_{\beta}^{(\psi)2}}
  {m_{\beta}^{(\psi)2}-M_{\alpha}^{(\psi)2}}
\ln \frac{m_{\beta}^{(\psi)2}}{M_{\alpha}^{(\psi)2}},
\end{eqnarray}
where $M_{\alpha}^{(\psi)}$ and $ m_{\beta}^{(\psi)}$ denote messenger
fermion masses and messenger sfermion masses respectively.  Under the
condition of Eq.(\ref{stable_cond}) the function ${\cal F}^{(\psi)}$
can be expanded in terms of
$F^{(\psi)}/(m_\psi m_{\bar{\psi}})$ as
\begin{eqnarray}
  {\cal F}^{(\psi)} =\frac{1}{2} \left| \frac{F^{(\psi)}}{m_\psi
    m_{\bar{\psi}}} \right|^2 \frac{F^{(\psi)}}{\sqrt{m_\psi
      m_{\bar{\psi}}}}
  {\cal{A}}^{(\psi)}(|V_{11}^{(\psi)}/V_{12}^{(\psi)}|^2,
|U_{11}^{(\psi)}/U_{12}^{(\psi)}|^2),
\label{approx_gaugino_mass}
\end{eqnarray}
where ${\cal{A}}^{(\psi)}(a,b)$ is
\begin{eqnarray}
  {\cal{A}}^{(\psi)}(a,b)
  &=&\frac{(ab)^{\frac{1}{4}}}{6(1-ab)^4(1+a)^{\frac{3}{2}}
    (1+b)^{\frac{3}{2}}} \left\{ 2(a+b)(-1+8ab-8 a^3 b^3 + a^4 b^4 +12
  a^2 b^2 \ln (ab)) \right.  \nonumber \\ && \left. -1-ab -64 a^2 b^2
  +64 a^3 b^3 + a^4 b^4 + a^5 b^5 -36 a^2 b^2 (1+ab) \ln (ab)
\right\}.
\label{A_function}
\end{eqnarray}
This function ${\cal{A}}^{(\psi)}(a,b)$ has the maximal value $0.1$ at
$a\simeq3$ and $b\simeq3$.  We should note that the leading term of
order $ F^{(\psi)}/\sqrt{m_\psi m_{\bar{\psi}}}$ vanishes because
$(M^{(\psi)})_{11}^{-1}=0$ ~\cite{INTY}.\footnote{This 
leading term cancellation of gaugino masses occurs generically, 
when we stabilize the SUSY breaking vacuum by mass terms 
as in Eq.(\ref{one_singlet_superpotential}).}

In the ordinary gauge mediation
models, contributions from the messenger quark multiplets
are equal to those from the messenger lepton multiplets
in the leading order of $ F/m$, and hence the GUT
relation among the gaugino masses,
$m_{\tilde{g}_1}/\alpha_1=m_{\tilde{g}_2}/\alpha_2=m_{\tilde{g}_3}/\alpha_3$,
holds.  In the present model, however, 
${\cal F}^{(d)} \neq {\cal F}^{(l)}$ 
because the leading term of order 
$ F^{(\psi)}/\sqrt{m_\psi m_{\bar{\psi}}}$ is canceled out.
Therefore, even if the unification of 
the Yukawa couplings and mass parameters is
assumed at the GUT scale, 
the GUT relation does not hold in
general.  From Eq.(\ref{gaugino_mass}) the following relation among
the gaugino masses
are satisfied:
\begin{eqnarray}
\frac{m_{\tilde{g}_1}}{\alpha_1}=\frac{3}{5}\frac{m_{\tilde{g}_2}}{\alpha_2}
+\frac{2}{5}\frac{m_{\tilde{g}_3}}{\alpha_3}.
\label{gaugino_mass_relation}
\end{eqnarray}
This is a distinctive prediction in the present model.

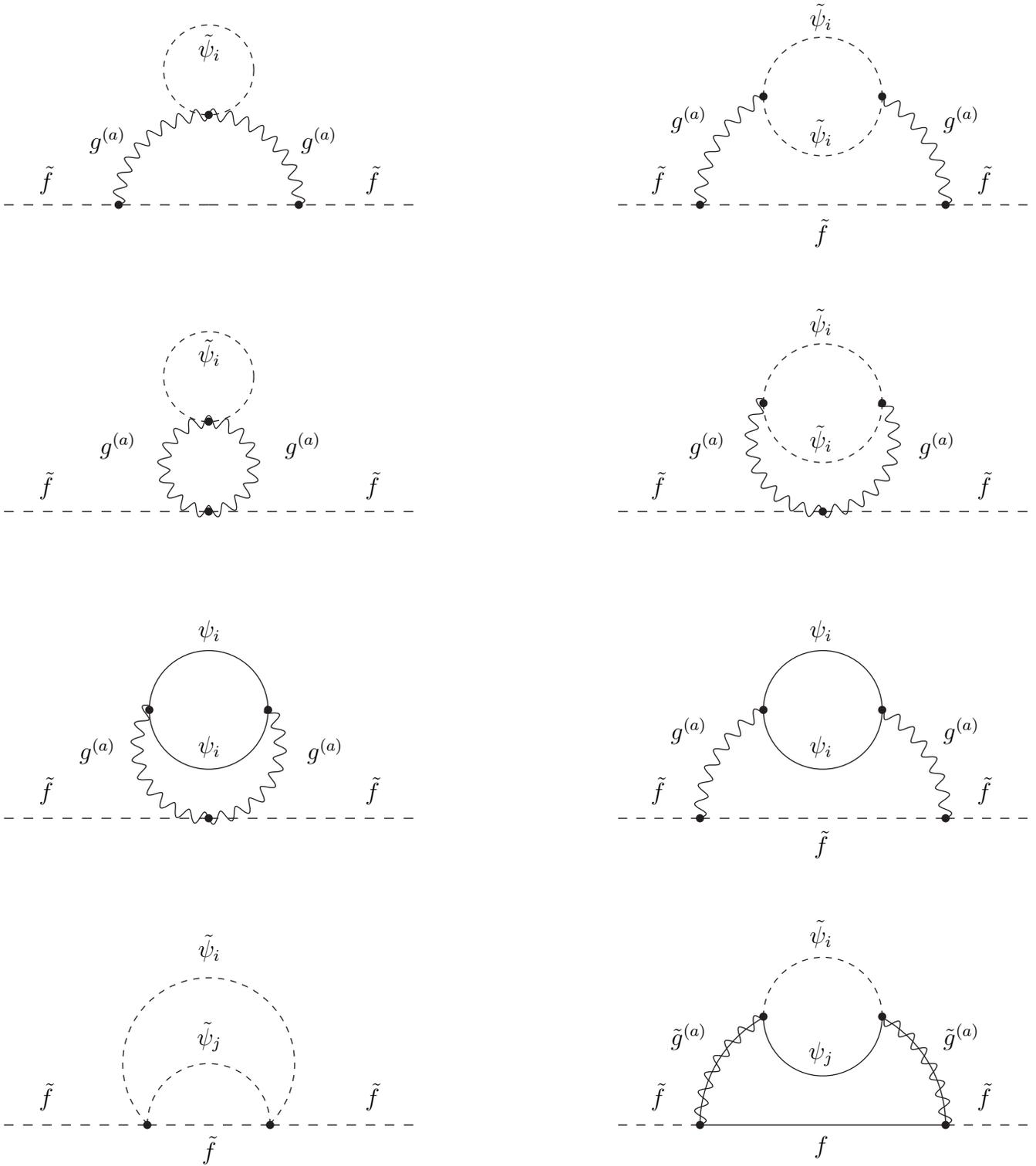
\begin{figure}
\begin{center} 
\begin{picture}(650,540)(-150,120)
% Graph 1
  \DashLine(-190,600)(-90,600){5} \Text(-170,610)[b]{$\tilde{f}$}
  \Vertex(-134,600){2} \Vertex(-46,600){2}
  \Text(-131,630)[rb]{$g^{(a)}$} \PhotonArc(-90,600)(44,0,180){3}{17}
  \Vertex(-90,644){2} \Text(-45,630)[lb]{$g^{(a)}$}
  \DashCArc(-90,666)(22,0,360){3} \Text(-90,675)[b]{$\tilde{\psi}_i$}
  \DashLine(-90,600)(10,600){5} \Text(-10,610)[b]{$\tilde{f}$}

% Graph 3
  \DashLine(-190,450)(-90,450){5} \Text(-170,460)[b]{$\tilde{f}$}
  \Vertex(-90,450){2} \Text(-126,480)[rb]{$g^{(a)}$}
  \PhotonArc(-90,472)(22,0,360){3}{17} \Vertex(-90,494){2}
  \Text(-53,480)[lb]{$g^{(a)}$} \DashCArc(-90,516)(22,0,360){3}
  \Text(-90,525)[b]{$\tilde{\psi}_i$} \DashLine(-90,450)(10,450){5}
  \Text(-10,460)[b]{$\tilde{f}$}

% Graph 5
  \DashLine(-190,300)(-90,300){5} \Text(-170,310)[b]{$\tilde{f}$}
  \Vertex(-90,300){2} \PhotonArc(-90,335)(35,145,35){3}{18}
  \Text(-136,330)[rb]{$g^{(a)}$} \Vertex(-119,353){2}
  \Text(-42,330)[lb]{$g^{(a)}$} \Vertex(-61,353){2}
  \CArc(-90,353)(29,0,180) \Text(-90,390)[b]{$\psi_i$}
  \CArc(-90,353)(29,180,360) \Text(-90,333)[b]{$\psi_i$}
  \DashLine(-90,300)(10,300){5} \Text(-10,310)[b]{$\tilde{f}$}

% Graph 7
  \DashLine(-190,150)(-120,150){5} \Text(-170,160)[b]{$\tilde{f}$}
  \Vertex(-120,150){2} \DashLine(-120,150)(-60,150){5}
  \Text(-90,145)[t]{$\tilde{f}$} \DashCArc(-90,150)(30,0,180){3}
  \DashCArc(-90,180)(42,315,225){3}
  \Text(-90,233)[b]{$\tilde{\psi}_i$}
  \Text(-90,190)[b]{$\tilde{\psi}_j$} \DashLine(-60,150)(10,150){5}
  \Text(-10,160)[b]{$\tilde{f}$} \Vertex(-60,150){2}

% Graph 2
  \DashLine(110,600)(150,600){5} \Text(130,610)[b]{$\tilde{f}$}
  \Vertex(150,600){2} \DashLine(150,600)(270,600){5}
  \Text(210,595)[t]{$\tilde{f}$} \Text(154,640)[rb]{$g^{(a)}$}
  \PhotonArc(210,600)(60,120,180){3}{7} \Vertex(181,653){2}
  \Text(270,640)[lb]{$g^{(a)}$} \PhotonArc(210,600)(60,0,60){3}{7}
  \Vertex(239,653){2} \DashCArc(210,653)(29,0,180){3}
  \Text(210,690)[b]{$\tilde{\psi}_i$}
  \DashCArc(210,653)(29,180,360){3}
  \Text(210,633)[b]{$\tilde{\psi}_i$} \DashLine(270,600)(310,600){5}
  \Text(290,610)[b]{$\tilde{f}$} \Vertex(270,600){2}

% Graph 4
  \DashLine(110,450)(210,450){5} \Text(130,460)[b]{$\tilde{f}$}
  \Vertex(210,450){2} \PhotonArc(210,485)(35,145,35){3}{18}
  \Text(163,480)[rb]{$g^{(a)}$} \Vertex(181,503){2}
  \Text(258,480)[lb]{$g^{(a)}$} \Vertex(239,503){2}
  \DashCArc(210,503)(29,0,180){3} \Text(210,540)[b]{$\tilde{\psi}_i$}
  \DashCArc(210,503)(29,180,360){3}
  \Text(210,483)[b]{$\tilde{\psi}_i$} \DashLine(210,450)(310,450){5}
  \Text(290,460)[b]{$\tilde{f}$}

% Graph 6
  \DashLine(110,300)(150,300){5} \Text(130,310)[b]{$\tilde{f}$}
  \Vertex(150,300){2} \DashLine(150,300)(270,300){5}
  \Text(210,295)[t]{$\tilde{f}$} \Text(154,340)[rb]{$g^{(a)}$}
  \PhotonArc(210,300)(60,120,180){3}{7} \Vertex(181,353){2}
  \Text(270,340)[lb]{$g^{(a)}$} \PhotonArc(210,300)(60,0,60){3}{7}
  \Vertex(239,353){2} \CArc(210,353)(29,0,180)
  \Text(210,390)[b]{$\psi_i$} \CArc(210,353)(29,180,360)
  \Text(210,333)[b]{$\psi_i$} \DashLine(270,300)(310,300){5}
  \Text(290,310)[b]{$\tilde{f}$} \Vertex(270,300){2}

% Graph 8
  \DashLine(110,150)(150,150){5} \Text(130,160)[b]{$\tilde{f}$}
  \Vertex(150,150){2} \Line(150,150)(270,150) \Text(210,145)[t]{$f$}
  \CArc(210,150)(60,120,180) \Text(154,190)[rb]{$\tilde{g}^{(a)}$}
  \PhotonArc(210,150)(60,120,180){3}{7} \Vertex(181,203){2}
  \CArc(210,150)(60,0,60) \Text(270,190)[lb]{$\tilde{g}^{(a)}$}
  \PhotonArc(210,150)(60,0,60){3}{7} \Vertex(239,203){2}
  \DashCArc(210,203)(29,0,180){3} \Text(210,240)[b]{$\tilde{\psi}_i$}
  \CArc(210,203)(29,180,360) \Text(210,183)[b]{$\psi_j$}
  \DashLine(270,150)(310,150){5} \Text(290,160)[b]{$\tilde{f}$}
  \Vertex(270,150){2}

%\put(37,80){\Huge{Fig.2}}
\end{picture}
\caption{Diagrams contributing to the squark, slepton and Higgs masses} 
\label{sfermion_mass_graph}
\end{center}
\end{figure}

The soft SUSY breaking masses for squarks, sleptons, and Higgses
$\tilde{f}$ in the MSSM sector are generated by two-loop diagrams
shown in Fig.\ref{sfermion_mass_graph} \cite{DNS,Martin,DG}.
We obtain them as
\begin{eqnarray}
  m^2_{\tilde{f}}=\frac{1}{2} \left[ C_3^{\tilde{f}}
  \left(\frac{\alpha_3}{4 \pi} \right)^2 {\cal G}^{(d)2} +
    C_2^{\tilde{f}} \left(\frac{\alpha_2}{4 \pi} \right)^2 {\cal
      G}^{(l)2} + \frac{3}{5} Y^2 \left(\frac{\alpha_1}{4 \pi}
  \right)^2 \left(\frac{2}{5} {\cal G}^{(d)2}
  + \frac{3}{5} {\cal G}^{(l)2} \right) \right],
\label{sfermion_mass}
\end{eqnarray}
where $C_3^{\tilde{f}}=\frac{4}{3}$ and $C_2^{\tilde{f}}=\frac{3}{4}$
when $\tilde{f}$ is in the fundamental representation of SU(3)$_C$ and
SU(2)$_L$, and $C_i^{\tilde{f}}=0$ for the gauge singlets, and $Y$
denotes the U(1)$_Y$ hypercharge ($Y \equiv Q-T_3$). Here ${\cal
  G}^{(d)2}$ and ${\cal G}^{(l)2}$ represent the contributions from
the messenger quark and lepton multiplets,
respectively, and they are given in Appendix A in details.
In contrast to the gaugino masses, 
the leading term of order $ F^{(\psi)}/\sqrt{m_\psi
  m_{\bar{\psi}}}$ is not canceled out for the
sfermion masses. Therefore, the gaugino masses
have an extra suppression factor $|F^{(\psi)}/(m_\psi
m_{\bar{\psi}})|^2$ compared with the sfermion masses. The 
lighter gauginos are a typical feature of this model.

Since the global SUSY is spontaneously broken, there is 
a Nambu-Goldstone fermion (goldstino) for the SUSY breaking. In
the framework of local SUSY (supergravity), the goldstino becomes
the longitudinal component of the gravitino.
Then, the gravitino has a mass given by 
\begin{eqnarray}
  m_{3/2} &=& \frac{F_{Z_1}+F_{Z_2}+F_{Z_3}}{\sqrt{3}M_{*}}, \nonumber
  \\ &=& \left( \frac{\sqrt{F_{Z_1}+F_{Z_2}+F_{Z_3}}}{2\times
    10^6~\GEV}
\right)^2~\KEV.
\end{eqnarray}
Here, we have imposed the vanishing cosmological constant and
$M_{*}=2.4\times 10^{18}$ GeV is the reduced Planck mass. If the SUSY
breaking scale is smaller than $2 \times 10^6~\GEV$, the gravitino mass is
smaller than $1~\KEV$ and hence the cosmological requirement is
satisfied.  
If all Yukawa couplings $k_{i}, \lambda_{i}, f_{i}, f_{\psi}, f_{\bar{\psi}}$
are $O(1)$ the squark mass $m_{\tilde{q}}$, for example, is 
roughly given by 
\begin{eqnarray}
  m_{\tilde{q}} \sim \frac{\alpha_{3}}{4\pi}\sqrt{F_{Z}}.
\end{eqnarray}
We expect $\sqrt{F_{Z}} \sim O(10^5)~\GEV$ to obtain 
$m_{\tilde{q}} \sim O(10^3)~\GEV$.
Thus $m_{3/2} < 1~\KEV$ is a natural result in the present model.

The gaugino and sfermion masses in Eqs.(\ref{gaugino_mass},
\ref{sfermion_mass}) are given only at the messenger scale.
Therefore, we must reevaluate them at the electroweak scale by using the
renormalization group equations (RGEs).  Here we present numerical
results of the mass spectrum of the gauginos and sfermions
including the running effects to the electroweak scale 
from the messenger scale.  To see the dependence of 
$|F^{(\psi)}/m_\psi m_{\bar{\psi}}|$, we
assume the following mass parameter relation in the messenger
sector Eq.(\ref{mass_matrix}) for simplicity:
\begin{eqnarray}
  m^{(\psi)}=m_\psi=m_{\bar{\psi}} \equiv \Lambda^{(\psi)}
\label{mass_assumption1}
\end{eqnarray}
for $\psi=d,l$.  The mass spectrum of the superparticles
in the MSSM sector is shown as a function of $|F^{(\psi)}/m_\psi
m_{\bar{\psi}}|$ in Fig.{\ref{sparticle_mass_figure}. Here we have set the
  scales
$\Lambda^{(\psi)}$ for $\psi=d,l$ as 
\begin{eqnarray}
\Lambda^{(d)}=\Lambda^{(l)}=5.0 \times~10^5~\GEV,
\end{eqnarray}
which corresponds to the gravitino mass $m_{3/2}\sim 0.1~\KEV$
for $\lambda_1=\lambda_2=\lambda_3=f_1=f_3=f_\psi=f_{\bar{\psi}}=1$.
\begin{figure}
  \centerline{ \psfig{file=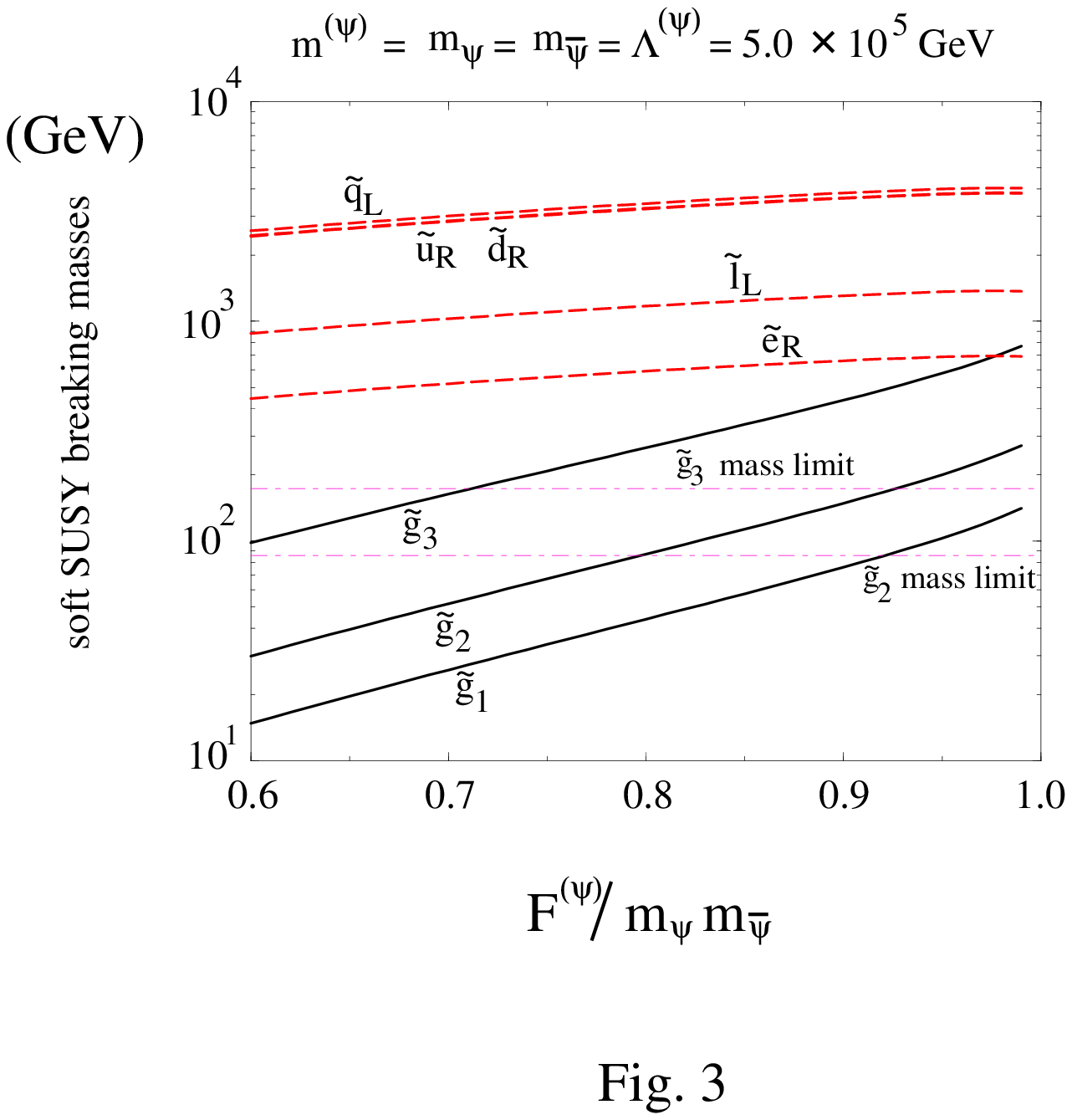,width=15cm}}
\caption{The mass spectrum of the gauginos and sfermions 
  as a function of the parameter $|F^{(\psi)}/m_\psi
  m_{\bar{\psi}}|$. Here we have assumed the mass parameters in the
  messenger sector as $m^{(\psi)}=m_\psi=m_{\bar{\psi}} \equiv
  \Lambda^{(\psi)}$ for $\psi=d,l$. We have set the scales
  $\Lambda^{(d)}=\Lambda^{(l)}=5.0 \times~10^5~\GEV$. Solid lines
  represent the gluino $\tilde{g}_3$, wino $\tilde{g}_2$, and
  bino $\tilde{g}_1$ masses, the dashed lines denote the doublet
  squark $\tilde{q}_L$, right-handed sup $\tilde{u}_R$,
  right-handed sdown $\tilde{d}_R$, doublet slepton
  $\tilde{l}_L$, and right-handed selectron $\tilde{e}_R$. 
  The renormalization effects from the messenger scale 
  to the electroweak scale have been taken into account. We also
  show the experimental lower bounds on the gluino and wino masses
  (dash-dotted lines).}
\label{sparticle_mass_figure}
\end{figure}
As one can see from Fig.{\ref{sparticle_mass_figure}}, the gaugino
masses rapidly decrease as the parameter $|F^{(\psi)}/m_\psi
m_{\bar{\psi}}|$ gets smaller because there is the suppression factor
$|F^{(\psi)}/m_\psi m_{\bar{\psi}}|^2$ in the gaugino masses in
Eq.(\ref{approx_gaugino_mass}).  On the other hand, the sfermion
masses weakly depend on $|F^{(\psi)}/m_\psi m_{\bar{\psi}}|$
since the leading term of order $F^{(\psi)}/\sqrt{m_\psi
  m_{\bar{\psi}}}$ does not vanish in contrast to the gaugino masses.
Moreover, even if the parameter $|F^{(\psi)}/m_\psi m_{\bar{\psi}}|$
gets close to $1$, the gluino remains lighter than the squarks. The
reason is that there is a further suppression factor in 
a function ${\cal A^{(\psi)}}$ in Eq.(\ref{A_function}) 
of the mixing angles of the messenger mass matrices.  Thus, the
gauginos are always lighter than sfermions in the present 
model.\footnote{The mass spectrum that the squarks are relatively
  heavy compared with the wino is desirable from the viewpoint of the
  proton decay in the GUT case.  The experimental
  constraint from the proton decay is about
  $m_{\tilde{q}}^{2}/m_{\tilde{g_{2}}}~\gsim~10^{4}$ GeV
  in the relevant region \cite{Hisano}.}

Since the gaugino masses are much smaller than the sfermion masses,
the constraints on this model come from the experimental bounds on the
gaugino masses. Thus we also show the experimental lower bounds on the
gluino mass $(m_{\tilde{g}_3} > 173~\GEV)$ \cite{CDF} and the wino mass
$(m_{\tilde{g}_2} > 86~\GEV)$ \cite{LEP} in
Fig.\ref{sparticle_mass_figure}.\footnote{
The gluino mass is constrained as $m_{\tilde{g}_3} > 173~\GEV$
for large squark masses by the Tevatron experiment \cite{CDF}.
The lightest chargino mass is constrained as $m_{\tilde{\chi}_1^+}>86~\GEV$
for large sneutrino mass and large $\mu$-parameter by the LEP experiment
at $\sqrt{s}=172 \GEV$ \cite{LEP}. In our model, the squark and slepton 
masses are much larger than the gaugino masses. The $\mu$-parameter 
also tends to be large as we will discuss in section \ref{RAD_section}. 
Therefore, we use the gluino mass bound $m_{\tilde{g}_3} > 173~\GEV$ and
the wino mass bound $m_{\tilde{g}_2}> 86~\GEV$ here.}
The parameters $|F^{(d)}/m_d
m_{\bar{d}}|$ and $|F^{(l)}/m_l m_{\bar{l}}|$ are independent from each
other in general. Thus, they are constrained by the experimental bounds
on the gluino and wino masses independently.  From 
Fig.{\ref{sparticle_mass_figure}}, we see the parameters
$|F^{(d)}/m_d m_{\bar{d}}|$ and
$|F^{(l)}/m_l m_{\bar{l}}|$ are constrained as
\begin{eqnarray}
  0.71 &<& \left|\frac{F^{(d)}}{m_d m_{\bar{d}}} \right| <1, \nonumber
  \\ 
0.79 &<& \left|\frac{F^{(l)}}{m_l m_{\bar{l}}} \right| <1.
\label{constraint_fterm}
\end{eqnarray}
Here, the
upper bounds come from the vacuum stability condition
Eq.(\ref{stable_cond}).  As we discussed in the
previous section, the masses $m_{\psi}$ and
$m_{\bar{\psi}}$ originated from the same dynamics
as SUSY breaking.  Therefore it is natural that 
the parameters $|F^{(\psi)}/m_\psi m_{\bar{\psi}}|$ for $\psi=d,l$
are close to $1$.

One can see that the so-called GUT relation among the gaugino masses
does not hold since the parameters $|F^{(d)}/m_d
m_{\bar{d}}|$ and $|F^{(l)}/m_l m_{\bar{l}}|$ are independent from each
other. For example, when $|F^{(d)}/m_d m_{\bar{d}}|=0.75$ and
$|F^{(l)}/m_l m_{\bar{l}}|=0.95$, the gluino and wino 
have almost the same mass : $m_{\tilde{g}_3} \simeq m_{\tilde{g}_2}
\simeq210~\GEV$ as seen in Fig.\ref{sparticle_mass_figure}. 
It is remarkable that even in
this case the gaugino mass relation 
Eq.(\ref{gaugino_mass_relation}) holds and hence the bino mass is
determined by the other gaugino masses. 
Since the sfermion masses weakly change with
$|F^{(\psi)}/m_\psi m_{\bar{\psi}}|$, they are almost the same as seen in 
Fig.\ref{sparticle_mass_figure}.

In this analysis, we can not determine the $\mu$-parameter unless both the 
messenger quark and lepton mass parameters are fixed. However, since
the scalar masses are much larger than the gaugino masses, 
the $\mu$-parameter tends to be large ($\sim 1~\TEV$). 
Therefore the lightest neutralino and chargino are almost gauginos.
In particular, the lightest neutralino is mostly bino 
because the bino mass is smaller than the wino mass in the relevant 
parameter space.

So far we have assumed the mass parameter relation 
Eq.(\ref{mass_assumption1}). Next we discuss $m^{(\psi)}$ dependence.
We assume that $m_{\psi}=m_{\bar{\psi}}=\Lambda^{(\psi)}$ for
simplicity, and fix the parameter $|F^{(\psi)}/m_\psi
m_{\bar{\psi}}|=0.9$ and $\Lambda^{(\psi)}=5 \times
10^5~\GEV$ for $\psi=d,l$. 
Then the parameter $m^{(\psi)}$ dependence on the mass
spectrum of the superparticles in the MSSM sector is shown in
Fig.\ref{Z_dependence}. From the experimental bounds on the gluino and
wino masses, the parameters
$m^{(\psi)}/\sqrt{m_\psi m_{\bar{\psi}}}$ for $\psi=d,l$ are constrained as
\begin{eqnarray}
  0.21 &<& \frac{m^{(d)}}{\sqrt{m_d m_{\bar{d}}}} < 4.8, \nonumber \\ 
  0.36 &<& \frac{m^{(l)}}{\sqrt{m_l m_{\bar{l}}}} < 3.0.
\label{constraint_Z}
\end{eqnarray}
Since the parameters $m^{(\psi)}$ come from the vacuum expectation
values of the singlet fields $k_{\psi i} Z_i$ 
(See Eq.(\ref{mass_to_Z})), 
the above constraints require that the
scales $k_{\psi i} \langle Z_i \rangle$ are 
almost at the same order as the strong SU(2)
dynamical scale $\Lambda$.

\begin{figure}
  \centerline{
\psfig{file=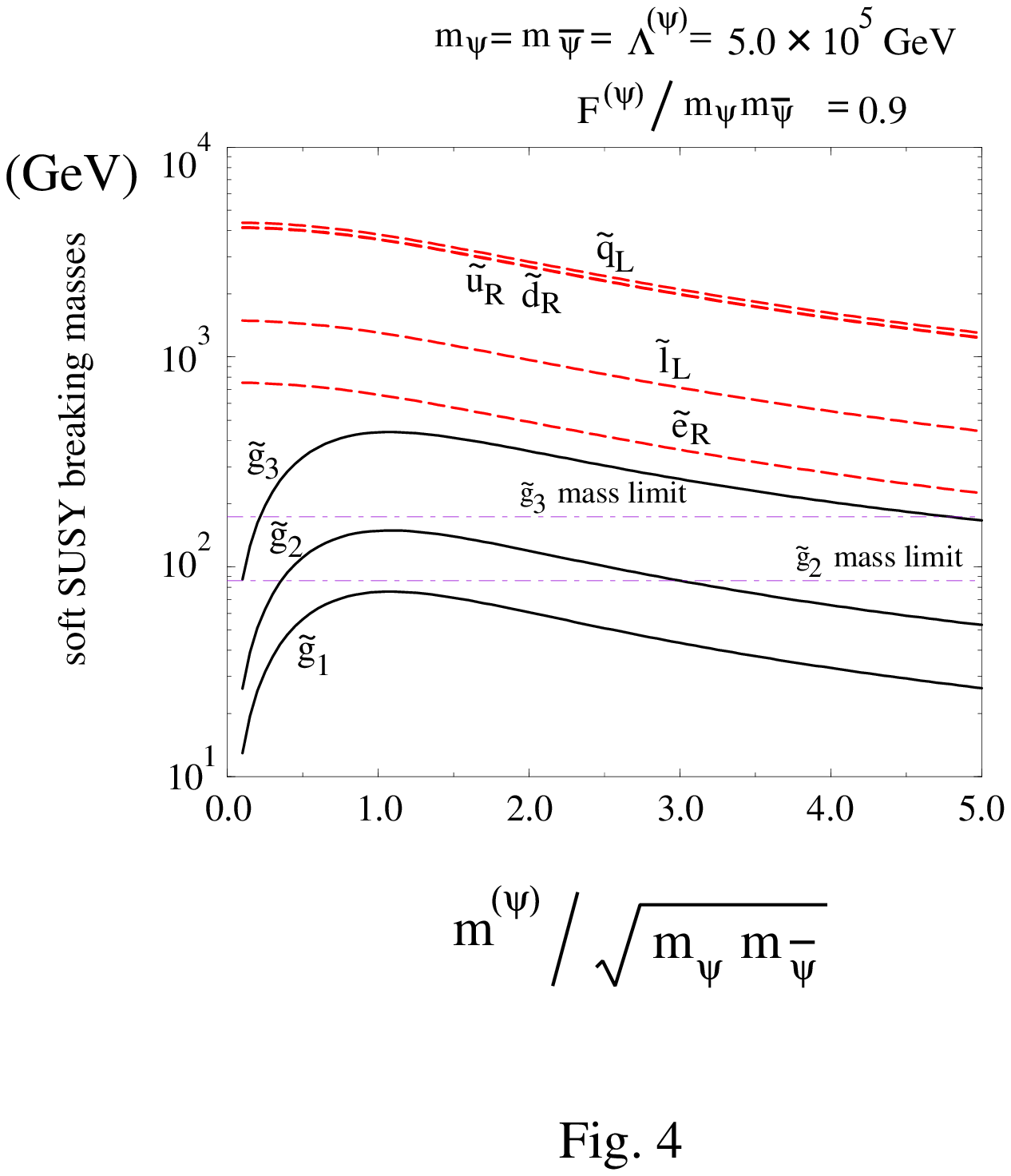,width=15cm}}
\caption{The mass spectrum of the gauginos and sfermions in the
  MSSM sector as a function of the parameter $m^{(\psi)}/\sqrt{m_\psi
    m_{\bar{\psi}}}$ for $\psi=d,l$. Here we have assumed the mass
  parameters in the messenger sector as $m_\psi=m_{\bar{\psi}} \equiv
  \Lambda^{(\psi)}$ for $\psi=d,l$. We have set the scales
  $\Lambda^{(d)}=\Lambda^{(l)}=5.0 \times~10^5~\GEV$. Solid lines
  represent the gluino $\tilde{g}_3$, wino $\tilde{g}_2$, and
  bino $\tilde{g}_1$ masses, the dashed lines denote the doublet
  squark $\tilde{q}_L$, right-handed sup $\tilde{u}_R$,
  right-handed sdown $\tilde{d}_R$, doublet slepton
  $\tilde{l}_L$, and right-handed selectron $\tilde{e}_R$. 
  The renormalization effects from the messenger scale 
  to the electroweak scale have been taken into account. We also
  show the experimental lower bounds on the gluino
  and wino masses.}
\label{Z_dependence}
\end{figure}
The constraints in Eq.(\ref{constraint_fterm}, \ref{constraint_Z})
will be weaken if the scales $\Lambda^{(\psi)}$ becomes larger. When we
fix the parameter $|F^{(d)}/m_d m_{\bar{d}}|$ ($|F^{(l)}/m_l
m_{\bar{l}}|$) and increase the scale $\Lambda^{(d)}$
($\Lambda^{(l)}$), then the gluino mass (the wino mass) becomes larger
proportionally to the scale $\Lambda^{(d)}$ ($\Lambda^{(l)}$) and the
function ${\cal{G}}^{(d)2}$ (${\cal{G}}^{(l)2}$) in the sfermion mass
squared also gets larger proportionally to $\Lambda^{(d)2}$
($\Lambda^{(l)2}$).  However, since the gravitino mass increases
proportionally to $\Lambda^{(\psi)2}$, too large value
$\Lambda^{(\psi)}$ conflicts with the cosmological bound on the
gravitino mass $m_{3/2} < 1~\KEV$. If the Yukawa couplings are set as
$\lambda_1=\lambda_2=\lambda_3=f_1=f_3=f_\psi=f_{\bar{\psi}}=1$ and we
take the mass parameter relation Eq.(\ref{mass_assumption1}), the
cosmological bound $m_{3/2} < 1~\KEV$ corresponds to
$\Lambda^{(d)}=\Lambda^{(l)} \leq 2\times~10^6~\GEV$.  When we set
$\Lambda^{(d)}=\Lambda^{(l)}=2\times10^6~\GEV$, the constraints on the
parameters $|F^{(d)}/m_d m_{\bar{d}}|$ and $|F^{(l)}/m_l m_{\bar{l}}|$
get weaker than those in Eq.(\ref{constraint_fterm}) as 
\begin{eqnarray}
  0.47 &<& \left| \frac{F^{(d)}}{m_d m_{\bar{d}}} \right| <1,
  \nonumber \\ 
0.54 &<& \left| \frac{F^{(l)}}{m_l m_{\bar{l}}} \right| <1.
\end{eqnarray}
The constraints on the parameters $|m^{(d)}/\sqrt{m_d
  m_{\bar{d}}}|$
and $|m^{(l)}/\sqrt{m_l m_{\bar{l}}}|$ are also much weaken,
\begin{eqnarray}
  0.05 &<& \frac{m^{(d)}}{\sqrt{m_d m_{\bar{d}}}} < 19,
  \nonumber \\ 
0.08  &<& \frac{m^{(l)}}{\sqrt{m_l m_{\bar{l}}}} < 12,
\end{eqnarray}
where we have taken the mass parameters
$m_{\psi}=m_{\bar{\psi}}=\Lambda^{(\psi)}=2\times 10^6~\GEV$ 
and $|F^{(\psi)}/m_\psi
m_{\bar{\psi}}|=0.9$.  In the case of the maximal scale
$\Lambda^{(\psi)} \simeq 2 \times 10^6~\GEV$, 
however, the squarks become much heavier than the 
electroweak scale ($m_{\tilde{q}} \simeq 10~\TEV$). 
Therefore, we need a fine tuning of $\mu$-parameter in
order to break the electroweak symmetry correctly as we will discuss
in the next subsection.

\subsection{Radiative electroweak symmetry breaking}
\label{RAD_section}

In the framework of the low-energy gauge-mediated SUSY
breaking scenario, the radiative electroweak symmetry breaking is
also realized as discussed in
Refs. \cite{Rad_sym_break_MMM1,Rad_sym_break_MMM2}.  In the model we
consider, the radiative electroweak symmetry breaking occurs
as in the ordinary gauge mediation models.

The soft SUSY breaking masses for the squarks, sleptons, and Higgses
are generated at the messenger scale as discussed in the previous
subsection.  When we include the running effects of the RGEs, the soft
SUSY breaking masses receive significant corrections from the
Yukawa interactions as well as the gauge interactions. The soft SUSY
breaking masses, especially for $H_2$ which is the MSSM Higgs doublet 
and couples to the top quark, receive significant corrections 
from the large top Yukawa interaction.  The approximate solution to
the RGE of the soft SUSY breaking mass squared
of $H_2$, $m^2_{H_2}$, is given by an iteration as follows:
\begin{eqnarray}
  m^2_{H_2} \simeq m^2_{H_1}-\frac{f_{t}^2}{16 \pi^2} 12
  m^2_{\tilde{t}}
\log \left( \frac{\Lambda_{mess}}{m_{\tilde{t}}} \right),
\label{RGE_solution_higgs}
\end{eqnarray}
where $m_{H_1}$ and $m_{\tilde{t}}$ are the soft SUSY breaking masses
for another Higgs doublet and the stop, respectively. $\Lambda_{mess}$
denotes the messenger scale, and $f_t$ represents the top Yukawa
coupling constant.  In the present model, the stop is heavier
than the SU(2) doublet Higgses because $\alpha_3>\alpha_2$.  Then
it drives $m_{H_2}^2$ to a negative value even if the running distance
between the messenger scale $\Lambda_{mess} \sim 10^5~\GEV$ and the
electroweak scale $m_{weak}\sim 10^2~\GEV$ is not so long.  Therefore, 
the electroweak symmetry is broken radiatively.

Requiring that the tree level potential has an extremum at vacuum
expectation values for the two Higgs doublets as $\langle H_1 \rangle
=v \cos \beta/\sqrt{2}$ and $ \langle H_2 \rangle= v \sin
\beta/\sqrt{2}$,
one finds
\begin{eqnarray}
  \frac{m^2_Z}{2} &=& \frac{m^2_{H_1}-m^2_{H_2} \tan^2 \beta}{\tan^2
    \beta-1} -\mu^2,
\label{det_mu}
\\
\sin 2\beta &=& - \frac{B \mu}{m^2_{H_1}+m^2_{H_2}+2 \mu^2}.
\label{det_b}
\end{eqnarray}

When we fix $\tan \beta$, the $\mu$-parameter is determined from
Eq.(\ref{det_mu}) to reproduce the correct value for the Z boson mass
$m_Z$.  Once the parameter $\mu$ is fixed, the
$B$-parameter (soft SUSY breaking mass for the Higgs doublet 
defined as ${\cal L} = B\mu H_{1} H_{2}$) 
is also determined from Eq.(\ref{det_b}).

Here we present numerical results of the radiative electroweak
symmetry breaking.  In our numerical analysis, we fix $\tan \beta$,
and we use the one-loop effective potential \cite{one_loop_eff_pot} to
determine the $\mu$- and $B$-parameters. We include all of the third
generation Yukawa couplings.  
Here, we assume that some unknown dynamics generates 
the $\mu$- and $B$- terms \cite{DDR}.
As for $A$-parameters, which are
trilinear couplings of scalars, we assume that 
they are very small at the messenger scale
$\Lambda$ : $A_{\tilde f}(\Lambda) \simeq 0$, because they are
generated only through higher loop diagrams in 
all known gauge mediation models. 
Under this initial
condition at the messenger scale, we solve the RGEs for the
$A$-parameters and calculate them at the electroweak scale to evaluate
the one-loop effective potential.

We show the $\mu$-parameters which satisfy the condition of
the radiative electroweak symmetry breaking as a function of the
parameter
$|F^{(\psi)}/m_\psi m_{\bar{\psi}}|$ in Fig.\ref{muterm_radiative}.
\begin{figure}
  \centerline{ \psfig{file=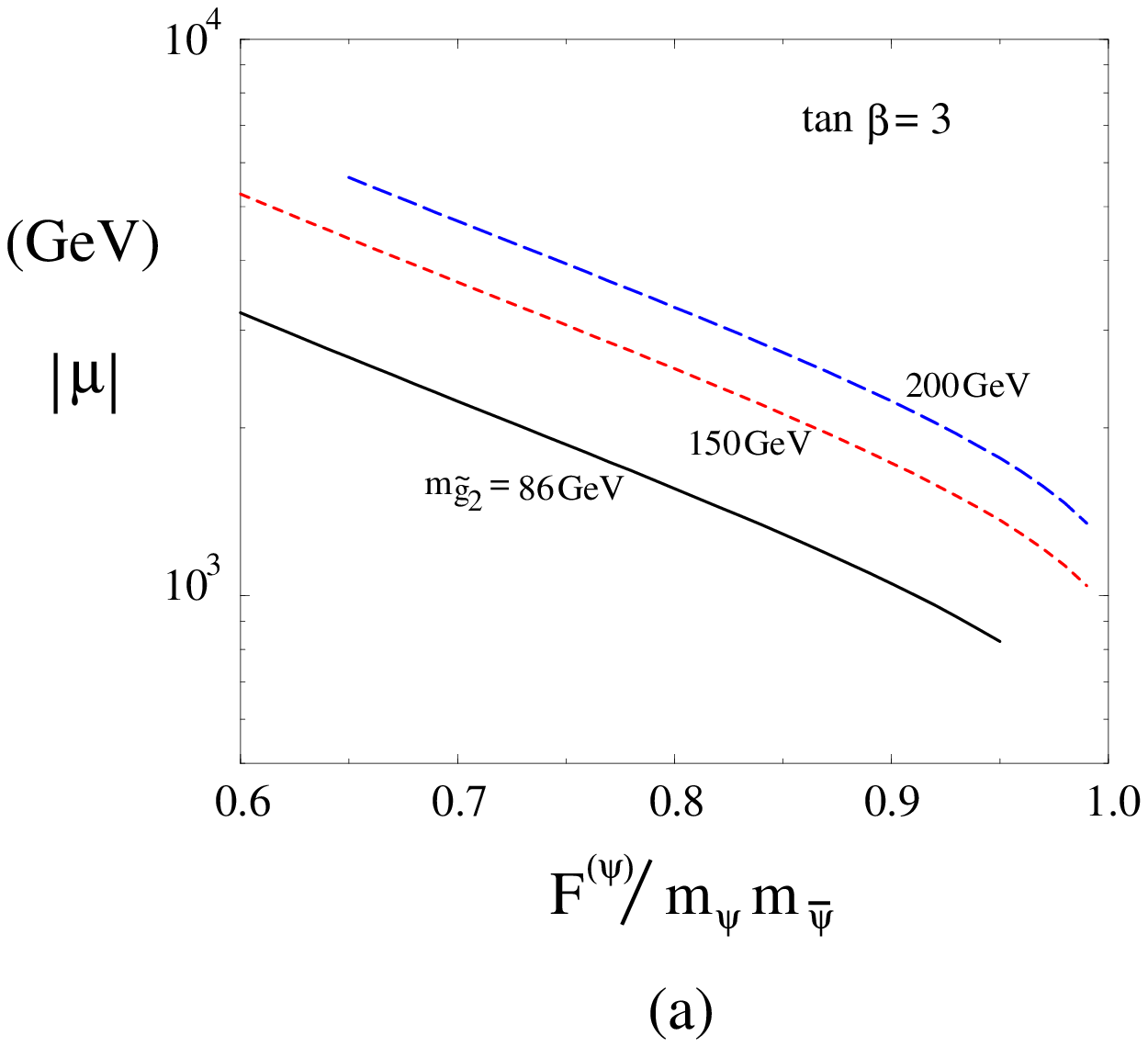,width=10cm}} \centerline{
    \psfig{file=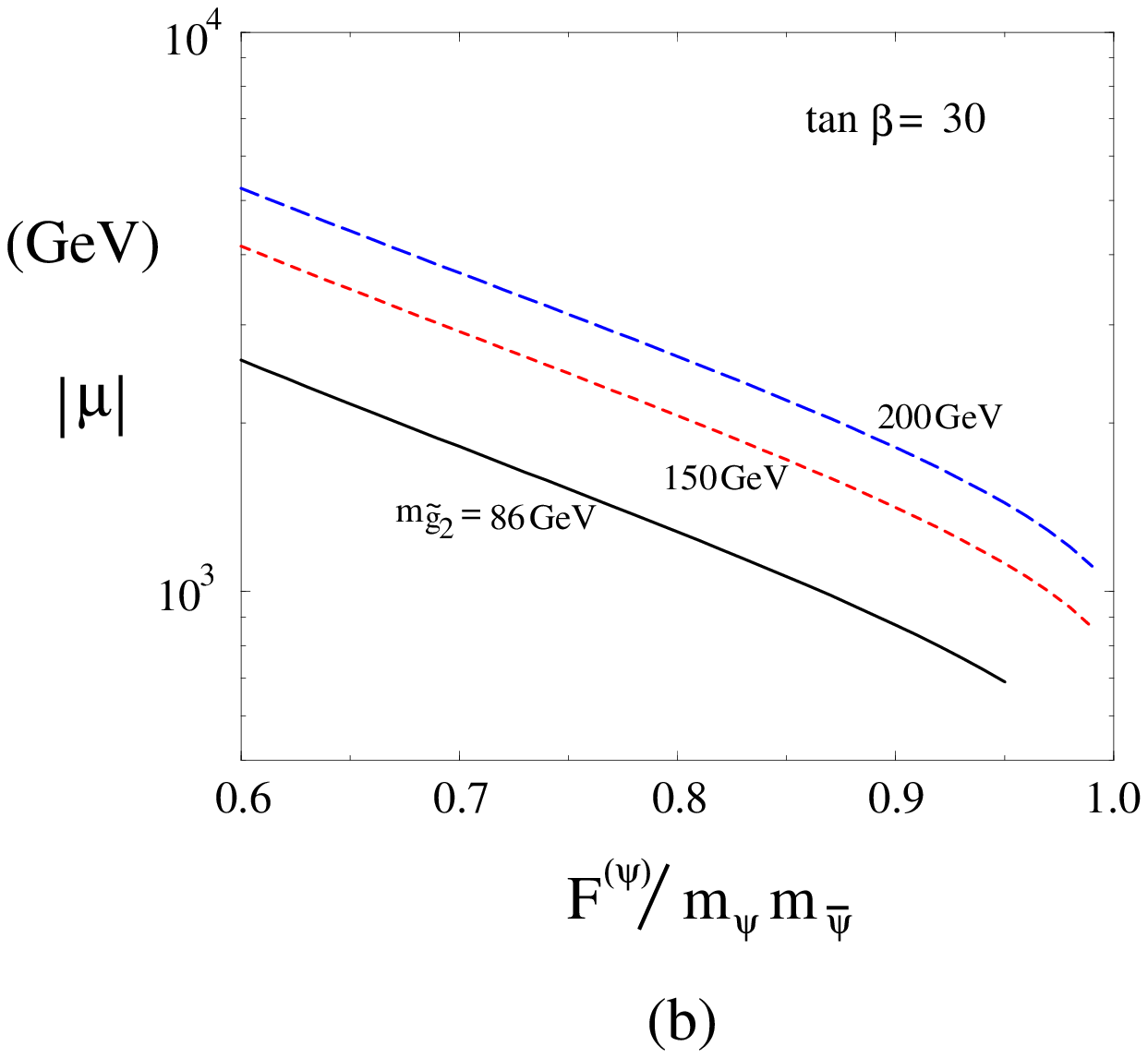,width=10cm}}
\caption{The $\mu$-parameters which satisfy the condition of
  the radiative electroweak symmetry breaking as a function of the
  parameter $F^{(\psi)}/m_\psi m_{\bar{\psi}}$.  Here we take (a)
  $\tan \beta=3$ and (b) $30$.  Solid line, dashed line, and
  long-dashed line correspond to the wino mass
  $m_{\tilde{g}_2}=86~\GEV$ (the experimental lower bound), 
  $150~\GEV$,   and $200~\GEV$, respectively.}
\label{muterm_radiative}
\end{figure}
Here we have assumed the mass parameter relation 
Eq.(\ref{mass_assumption1}) and $\Lambda^{(d)}=\Lambda^{(l)}$ 
in the messenger sector 
and we have taken $\tan \beta=3$
(Fig.\ref{muterm_radiative}(a)), $30$ (Fig.\ref{muterm_radiative}(b)).
Solid line, dashed line, and long-dashed line correspond to the wino
mass $m_{\tilde{g}_2}=86~\GEV$ (the experimental lower bound), $150~\GEV$, 
and $200~\GEV$, respectively.  As one sees from
Fig.\ref{muterm_radiative}, the $\mu$-parameter gets larger as the
parameter $|F^{(\psi)}/m_\psi m_{\bar{\psi}}|$ becomes smaller when we
fix the gaugino masses. The reason is as follows: if the parameter
$|F^{(\psi)}/m_\psi m_{\bar{\psi}}|$ is too small, the gauginos are
much lighter than their experimental mass bounds since the gaugino
masses have the suppression factor $|F^{(\psi)}/m_\psi
m_{\bar{\psi}}|^2$. To exceed the experimental bound on the gaugino
masses, the messenger scale is required to be much larger. Then,
the soft SUSY breaking masses for the sfermions get larger since they
do not have the suppression like gauginos.  Since the squarks,
especially stop, become much heavier, the mass squared of $H_2$ becomes
too negative as one can see in Eq.(\ref{RGE_solution_higgs}). Then one
need a large $\mu$-parameter to
reproduce the correct value of the Z boson mass. (See Eq.(\ref{det_mu}).)
For example, in the case with $\tan \beta=3$ and $|F^{(\psi)}/m_\psi
m_{\bar{\psi}}|=0.6$, we need a large $\mu$-parameter as
$|\mu|~\gsim~3100~\GEV$. In this way, we may need a fine tuning of
$\mu$-parameter when the parameter $|F^{(\psi)}/m_\psi
m_{\bar{\psi}}|$ is small (i.e. the SUSY breaking scale is
large). Therefore, from such a naturalness point of view the light
gravitino $m_{3/2}\sim 0.01-1~\KEV$ is implied.
On the other hand, as
$\tan \beta$ becomes larger, the $\mu$-parameter gets smaller because
the mass squared of $H_2$ becomes less negative due to the relatively 
smaller top Yukawa coupling. 

%Since the $\mu$-parameter is much larger than the gaugino masses,
%the lightest neutralino and chargino consist of the bino and wino,
%on the other hand, the heavy neutralino and chargino are almost
%higgsinos in our model.
%
We finally comment on the $\mu$-term generation \cite{DNS,DGP,Yana}.
In the present model, the $\mu$-term may be generated in the same way as the
generation of the messenger mass parameters: if the superfield $X$
couples to $H_1 H_2$ , the SUSY invariant mass $\mu$ for Higgses $H_1$
and $H_2$ is generated. Since $\langle X \rangle \simeq
10^{5-6}~\GEV$, we need a small coupling constant
$\lambda_h \simeq
10^{-3}$, where $\lambda_h$ is defined by $W=\lambda_h X H_1 H_2$, to
have the desired value $\mu \simeq (10^2-10^3)~\GEV$. 
The small $\lambda_h$ is natural
in the sense of 't Hooft.  We note that no large $B$-term ($B\mu H_1
H_2$) is induced since the $F$-component of $X$ is very small.\footnote{
In this case, the SUSY CP problem is also solved. When we consider
the GUT models, the phases of the gaugino masses can be eliminated by
a common rotation of the gauginos since $k_{d2}/k_{d3} \simeq k_{l2}/k_{l3}$
holds even at the low-energy scales. Then the rotation of the gauginos gives
rise to a phase in the Yukawa-type gauge couplings of the gauginos.
Such a phase can be eliminated by a rotation of the sfermions and 
Higgses since there are no $A$-terms and no $B$-term at the tree level.}
Hence
the scale $\mu$ may originate from the same dynamics as the SUSY
breaking.  In this case, the large $\tan\beta \sim 50$ is required 
because of
the small $B$-parameter to break electroweak symmetry radiatively.

\section{The Grand Unified Model}
\subsection{The GUT model with the Yukawa unification}

The messenger quarks and leptons $(d,l)$, $(d',l')$,
$(\bar{d},\bar{l})$, and $(\bar{d}',\bar{l}')$ are embedded in $\bf 5$
and $ {\bf 5}^*$ representations of the SU(5) group. Therefore, if we
extend our model to the GUT model (without
non-renormalizable interactions), certain Yukawa coupling constants in
Eq.(\ref{superpotential_model3})
are unified as
\begin{eqnarray}
  k_{di} &=& k_{li}, ~~~~(i=1-3),
\label{unif_yukawa1}  \\
f_{d} &=& f_{l},
\label{unif_yukawa2}  \\
f_{\bar{d}} &=& f_{\bar{l}},
\label{unif_yukawa3}
\end{eqnarray}
at the GUT scale $M_{\rm GUT} \simeq 2 \times 10^{16}~\GEV$.  Under the
Yukawa unification Eqs.(\ref{unif_yukawa1}$-$\ref{unif_yukawa3}), we
can obtain the relations between the Yukawa couplings $k_{di}$ and
$k_{li}$, $f_d$ and $f_l$, and $f_{\bar{d}}$ and $f_{\bar{l}}$
at the messenger scale using the RGEs for these Yukawa coupling
constants.  We list the RGEs for the coupling constants in this model
in Appendix B.  In general, the gauge interactions increase the Yukawa
coupling constants at the lower scale. Since the messenger quarks have
the SU(3)$_C$ gauge interaction but the messenger leptons do not,
the Yukawa coupling constants $k_{di}$, $f_d$, and
$f_{\bar{d}}$ related to the messenger quarks tend to be larger than
the Yukawa couplings $k_{li}$, $f_l$, and $f_{\bar{l}}$ related to the
messenger leptons because $g_3(\mu) \ge g_2(\mu)$ at $\mu \le M_{\rm
  GUT}$.  We consider the RGEs for the ratios of the Yukawa
couplings $k_{di}/k_{li}$, $f_d/f_l$, and $f_{\bar{d}}/f_{\bar{l}}$;
\begin{eqnarray}
  16 \pi^2 \mu \frac{d}{d \mu}\left( \frac{k_{di}}{k_{li}} \right)
  &=&\frac{k_{di}}{k_{li}} \left\{ -\frac{16}{3} g_3^2 +3 g_2^2 +
  \frac{1}{3} g_1^2 \right.
\nonumber \\
&& \left.+2 \sum_{j=1}^{3}(k_{dj}^2-k_{lj}^2) +f_d^2 - f_l^2
+f_{\bar{d}}^2-f_{\bar{l}}^2 \right\}
\nonumber \\
&& +\sum_{j=1}^{3} \frac{k_{dj}}{k_{li}} \left(
1-\frac{k_{di}}{k_{li}} \frac{k_{lj}}{k_{dj}} \right) \left( \lambda_i
\lambda_j + 3 k_{di} k_{dj} +2 k_{li} k_{lj}
+2 f_i f_j \right),
\label{RGE_k_ratio}
\\
16 \pi^2 \mu \frac{d}{d \mu} \left( \frac{f_d}{f_l} \right) &=&
\frac{f_d}{f_l} \left\{ -\frac{16}{3} g_3^2 + 3 g_2^2 + \frac{1}{3}
g_1^2
+\sum_{i=1}^3 (k_{di}^2-k_{li}^2) + 2 (f_d^2 - f_l^2) \right\},
\label{RGE_f_ratio}
\\
16 \pi^2 \mu \frac{d}{d \mu} \left( \frac{f_{\bar d}}{f_{\bar l}}
\right) &=& \frac{f_{\bar d}}{f_{\bar l}} \left\{ -\frac{16}{3} g_3^2
+ 3 g_2^2 + \frac{1}{3} g_1^2
+\sum_{i=1}^3 (k_{di}^2-k_{li}^2) + 2 (f_{\bar d}^2 - f_{\bar l}^2) \right\}.
\label{RGE_barf_ratio}
\end{eqnarray}
One can see that under the condition Eqs.(\ref{unif_yukawa1}$-$\ref
{unif_yukawa3}) the ratios $k_{di}/k_{li}$, $f_d/f_l$, and
$f_{\bar{d}}/f_{\bar{l}}$ become larger at the lower scale 
because the SU(3)$_C$ gauge interaction
dominates over in
Eqs.(\ref{RGE_k_ratio}$-$\ref{RGE_barf_ratio}) as long as the
perturbative description of the Yukawa couplings is valid. 
When we impose the unification
condition Eqs.(\ref{unif_yukawa1}$-$\ref{unif_yukawa3}), the
right-hand side in Eq.(\ref{RGE_k_ratio}) is independent of the index
$i$. Therefore the ratio $k_{di}/k_{li}$ does not depend on the index
$i$ at any scales.  Then we can always set
$k_{d1}=k_{l1}=k_{d2}=k_{l2}=0$ and $f_2=0$ at any scales by an
unitary transformation of the singlet fields $Z_i$. Thus we work in
this basis below. Then the ratios
of the Yukawa couplings $k_{d3}/k_{l3}$, $f_d/f_l$, and
$f_{\bar{d}}/f_{\bar{l}}$ equal to those of the mass
parameters for the messenger sector $m^{(d)}/m^{(l)}$, $m_d/m_l$, and
$m_{\bar{d}}/m_{\bar{l}}$, respectively.  

To numerically analyze the relations between mass parameters in the
messenger sector, we fix some parameters\footnote{The Yukawa couplings
  $k_i, f_\psi, f_{\bar{\psi}}$ do not significantly affect the runnings
  of $\lambda_i$, $f_i$ as long as the perturbative description of the 
  Yukawa couplings is valid. Thus the vacuum expectation values 
  $\langle X \rangle$ and $\langle F_{Z_i} \rangle$ hardly depend on
  $k_i, f_\psi, f_{\bar{\psi}}$.} as $\lambda_Y=0.3$, $(\lambda_1, \lambda_2,
\lambda_3)=(0.2,0.1,0.16)$, $(f_1,f_2,f_3)=(0.5,0,0.6)$,
$(k_{d1},k_{d2})=(k_{l1},k_{l2})=(0,0)$ at the GUT scale,
$\Lambda=2\times 10^6~\GEV$ which is the dynamical scale of the strong
SU(2) gauge interaction, and $\langle Z_3 \rangle = 1\times 10^6~\GEV$
which is the vacuum expectation value of the singlet field $Z_3$. 
The strong SU(2) gauge coupling $g$ is taken as $2\pi$ at the messenger
scale (i.e. $\alpha=g^2/4\pi=\pi$). 
We solve the RGEs numerically varying 
$k_{d3}=k_{l3} \equiv k$, $f_d=f_l=f_\psi$, and
$f_{\bar{d}}=f_{\bar{l}}=f_{\bar{\psi}}$ at the GUT scale.
We find the ratios of the mass parameters given by
\begin{eqnarray}
  &&\frac{m^{(d)}}{m^{(l)}} \simeq 1.4~(1.4),
\label{mass_ratio_kd_kl}
\\
&&\frac{m_d}{m_l} =\frac{m_{\bar{d}}}{m_{\bar{l}}}\simeq 1.4~(1.4),
\label{mass_ratio_fd_fl}
\\
&&0.5~(0.4)< \frac{m^{(l)}}{\sqrt{m_l m_{\bar{l}}}}
\simeq  \frac{m^{(d)}}{\sqrt{m_d m_{\bar{d}}}}  < 1.2~(0.9),
\end{eqnarray}
for $0.42< k<1.0$ and $f_\psi=f_{\bar{\psi}}=0.20$ (for $0.32< k< 0.54$
and $f_\psi=f_{\bar{\psi}}=0.18$). Here we have taken
$f_\psi=f_{\bar{\psi}}$ for simplicity.  These parameters correspond
to the parameter region $F^{(l)}/m_l m_{\bar{l}}> 0.7$.  The ratio of
$m^{(d)}/m^{(l)}$ is almost the same as the ratio of
$m_d/m_l=m_{\bar{d}}/m_{\bar{l}}$ because the running effects for the
ratios of the Yukawa couplings $k_d/k_l$, $f_d/f_l$, and
$f_{\bar{d}}/f_{\bar{l}}$ dominantly come from the SU(3)$_C$ gauge
interaction as one can see from 
Eqs.(\ref{RGE_k_ratio}$-$\ref{RGE_barf_ratio}).  The parameter
$F^{(d)}/m_d
m_{\bar{d}}$ is also related to $F^{(l)}/m_l m_{\bar{l}}$ as follow:
\begin{eqnarray}
  \frac{F^{(l)}}{m_l m_{\bar{l}}} \simeq 1.4~\frac{F^{(d)}}{m_d
    m_{\bar{d}}},
\label{fl_fd_relation}
\end{eqnarray}
since 
\begin{eqnarray}
  \left(\frac{F^{(l)}}{m_l m_{\bar{l}}} \right) /
    \left(\frac{F^{(d)}}{m_d m_{\bar{d}}} \right) =
      \frac{k_{l3}}{k_{d3}} \frac{m_d}{m_l}
      \frac{m_{\bar{d}}}{m_{\bar{l}}}
=\frac{m^{(l)}}{m^{(d)}}\frac{m_d}{m_l}\frac{m_{\bar{d}}}{m_{\bar{l}}}.
\end{eqnarray}
Because of the Yukawa unification, $F^{(l)}/m_l m_{\bar{l}}$ becomes
larger than $F^{(d)}/m_d m_{\bar{d}}$ at the messenger scale.
This yields an interesting consequence on the mass
spectrum of the gauginos as we will see below.

We are now at the position to show the mass spectrum of the
gauginos and sfermions in the MSSM sector. 
\begin{figure}
\centerline{\psfig{file=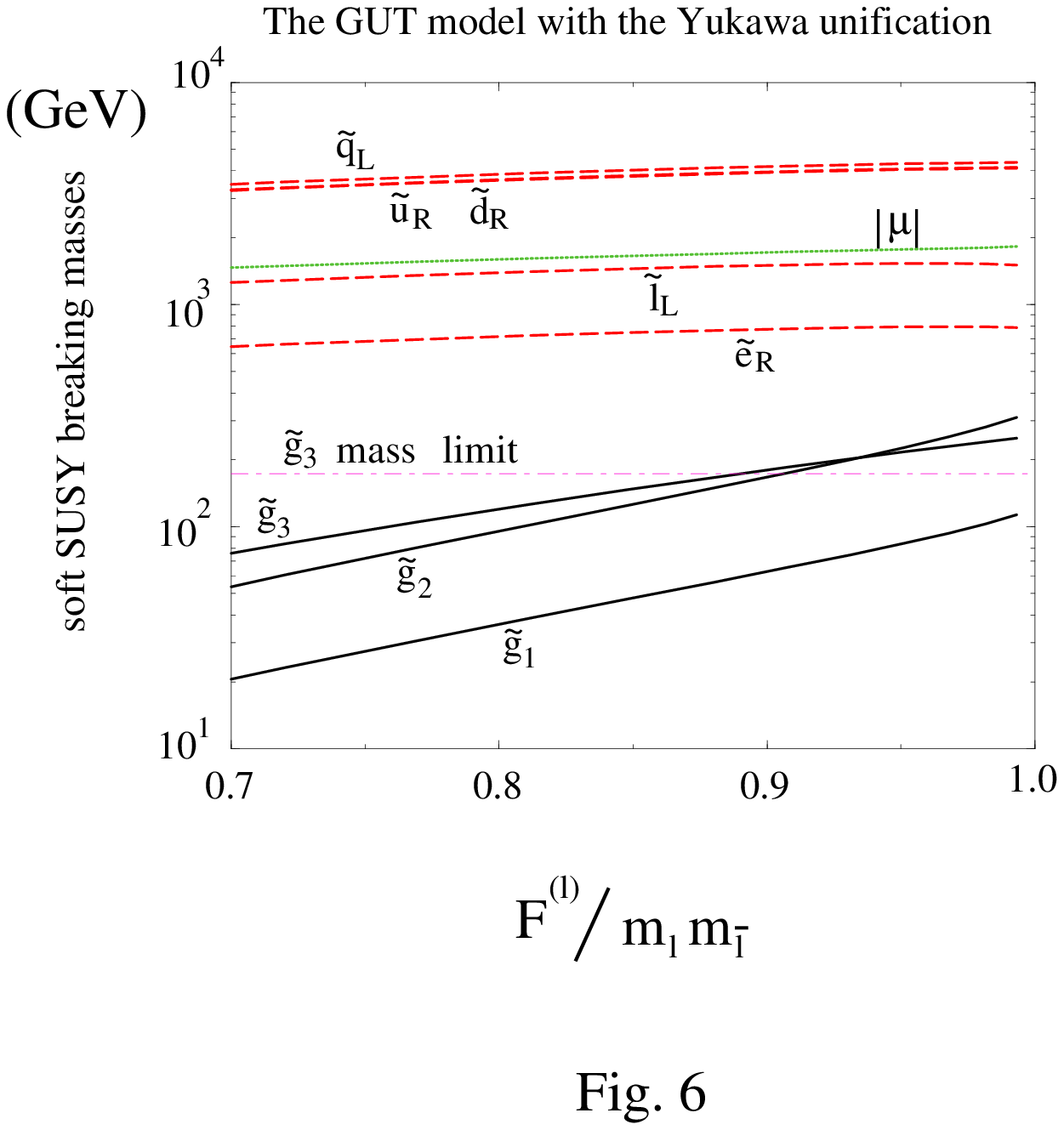,width=15cm}}
\caption{The mass spectrum of the gauginos and sfermions in the
  MSSM sector as a function of $F^{(l)}/m_l m_{\bar{l}}$. Here we have taken
  $\lambda_Y=0.3$, $(\lambda_1, \lambda_2,\lambda_3)=(0.2,0.1,0.16)$,
  $(f_1,f_2,f_3)=(0.5,0,0.6)$, $(k_{d1},k_{d2})=(k_{l1},k_{l2})=
  (0,0)$ at the GUT scale.  The dynamical scale $\Lambda$ of the
  strong SU(2) gauge interaction and the vacuum expectation value
  $\langle Z_3 \rangle$ have been taken as $\Lambda=2.0 \times 10^6~\GEV$
  and $\langle Z_3 \rangle=1.0 \times 10^6~\GEV$, respectively.
  We have set $f_{d}=f_{l}=f_{\bar{d}}=f_{\bar{l}} =0.2$ at the GUT scale
  for simplicity and varied $k_{l3}=k_{d3} \equiv k$ with $0.4< k <1.0$.
  Note that this parameter set corresponds to the gravitino mass
  $m_{3/2} \simeq 0.12~\KEV$.  Solid lines represent the gaugino
  masses: the bino $\tilde{g}_1$, wino $\tilde{g}_2$, and
  gluino $\tilde{g}_3$ masses, and dashed lines represent the sfermion
  masses: the left-handed squark $\tilde{q}_L$, right-handed
  sup $\tilde{u}_R$, right-handed sdown $\tilde{d}_R$, doublet
  slepton $\tilde{l}_L$, and
  right-handed selectron $\tilde{e}_R$ masses.
  The $\mu$-parameter is also shown (dotted line).
  The renormalization effects from the messenger scale 
  to the electroweak scale have been taken into account.
  We also show the experimental
  lower bound on the gluino mass (dash-dotted line).}
\label{mass_spectrum_GUT}
\end{figure}
In Fig.\ref{mass_spectrum_GUT}, the mass spectrum is shown as a
function of the parameter $F^{(l)}/m_l m_{\bar{l}}$.  Here we have taken the
same parameter set as above mentioned one and we have set
$f_\psi=f_{\bar{\psi}}=0.2$ and varied $k$ with $0.4<k<1$.  Note that
this parameter set corresponds to the gravitino mass $m_{3/2}
\simeq 0.12~\KEV$.  As
we have seen in the previous section, the gaugino masses strongly
depend on the parameters $F^{(\psi)}/m_\psi m_{\bar{\psi}}$ for
$\psi=d,l$ since they have an extra suppression factor
$(F^{(\psi)}/m_\psi m_{\bar{\psi}})^2$. When we impose the Yukawa
unification Eqs.(\ref{unif_yukawa1}$-$\ref{unif_yukawa3}), the
parameter $F^{(d)}/m_d m_{\bar{d}}$ is smaller than $F^{(l)}/m_l
m_{\bar{l}}$ as shown in Eq.({\ref{fl_fd_relation}}).  Therefore the
gluino mass receives a larger suppression than the wino mass, and hence the
gluino tends to be relatively light.  The experimental lower bound on the
gluino mass constrains the parameter $F^{(l)}/m_l m_{\bar{l}}$ to be 
$F^{(l)}/m_l m_{\bar{l}}> 0.89$. This leads to a constraint on the
Yukawa coupling $k$ at the GUT scale as $k>0.65$.\footnote{When we regard
  the masses for the gauginos and sfermions as a function of
  $F^{(l)}/m_l m_{\bar{l}}$, we obtain almost the same result as in
  Fig.\ref{mass_spectrum_GUT} even if we take a different value of the
  Yukawa coupling $f_\psi=f_{\bar{\psi}}$.  The reason is that the
  ratio between $F^{(l)}/m_l m_{\bar{l}}$ and $F^{(d)}/m_d
  m_{\bar{d}}$ is almost independent of $f_\psi=f_{\bar{\psi}}$.} Therefore,
the GUT relation among the gaugino masses does not hold
even though we consider the GUT model. We note that 
the gluino can be lighter than the wino.

We also show the $\mu$-parameter in Fig.\ref{mass_spectrum_GUT}.
As we discussed in section \ref{RAD_section}, the $\mu$-parameter
becomes much larger than the gaugino masses.

\subsection{The GUT model without the Yukawa unification}
So far we have considered the GUT model with the Yukawa unification.
In this section, we consider the GUT model with
non-renormalizable interactions.  The superpotential which contributes
to the Yukawa couplings $k_{\psi i}$,
$f_\psi$, and $f_{\bar{\psi}}$ $(\psi=d,l)$ at low energy scales is given by
\begin{eqnarray}
  W &=&\sum_{i=1}^3 Z_i \left ( k_{\psi_i}' \psi \bar{\psi}
  +k_{\psi_i}'' \frac{\Sigma}{M_*} \psi \bar{\psi} +\cdots \right)
  \nonumber \\ &&+X \left( f_\psi' \psi \bar{\psi'} + f_{\bar{\psi}}'
  \psi' \bar{\psi} +f_\psi'' \frac{\Sigma}{M_*} \psi \bar{\psi'}
  +f_{\bar{\psi}}'' \frac{\Sigma}{M_*} \psi' \bar{\psi}
\right),
\label{superpot_non_unif}
\end{eqnarray}
where the fields $\psi$ and $\psi'$ ($\bar{\psi}$ and $\bar{\psi'}$)
are $\bf{5}$ $(\bf{5^*})$ dimensional representation fields of SU(5)
group, which contain the messenger quark and lepton multiplets $(d,l)$ and
$(d',l')$ ( $(\bar{d},\bar{l})$ and $(\bar{d'}, \bar{l'})$ ),
respectively. The field $\Sigma$ is a $\bf{24}$ dimensional
representation which breaks SU(5) down to SU(3)$_C \times$SU(2)$_L \times$
U(1)$_Y$ with expectation value $\langle \Sigma \rangle =V {\rm
  diag}(2,2,2,-3,-3)$. We should notice that the
non-renormalizable interactions in Eq.(\ref{superpot_non_unif}) are
not forbidden by any symmetries provided that the $\Sigma$ is 
a trivial representation of U(1)$_R\times$ U(1)$_\chi$.  
Then, the Yukawa coupling constants
$k_\psi$, $f_\psi$, and $f_{\bar{\psi}}$
receive corrections of order of $O(\langle \Sigma \rangle /M_*)$ as
\begin{eqnarray}
  k_{di}&=&k'_{\psi i}+2k''_{\psi i} \frac{V}{M_*},
\\
k_{li}&=&k'_{\psi i}-3k''_{\psi i} \frac{V}{M_*},
\\
f_d &=& f_\psi' +2 f''_{\psi} \frac{V}{M_*},
\\
f_l &=& f_\psi' -3 f''_{\psi} \frac{V}{M_*},
\end{eqnarray}
at the GUT scale, and hence the Yukawa coupling unification of 
Eqs.(\ref{unif_yukawa1}$-$\ref{unif_yukawa3}) is broken in general.

In such a case, the relations between the messenger quark and lepton
mass parameters depend on the corrections from the non-renormalizable
terms, and the relation in Eq.(\ref{fl_fd_relation}) does not
hold. Thus the mass relation among the gauginos in the case with the
Yukawa unification may easily change.  Therefore, the existence of 
non-renormalizable terms weakens our prediction on the mass
spectrum of the superparticles.  However, it is important to study how
the mass spectrum in the previous subsection changes.

To demonstrate changes of the mass spectrum, we consider a simple example.  
Here we take the following
initial condition of the Yukawa couplings
at the GUT scale, 
\begin{eqnarray}
  k_{d2}=k_2-\delta,~&&~ k_{l2}=k_2,
\nonumber \\
k_{d3}= k_3-\delta,~&&~ k_{l3}= k_3,
\nonumber \\
f_d = f-\delta,~&&~ f_l = f,
\nonumber \\
f_{\bar{d}}=\bar{f}-\delta,~&&~
f_{\bar{l}}=\bar{f}.
\label{initial_nonunif}
\end{eqnarray}
The parameter $\delta$ represents the difference between the
Yukawa couplings of messenger quarks and leptons due to corrections
from non-renormalizable interactions. The case $\delta=0$
corresponds to the GUT model with the Yukawa unification.  We show the
mass spectrum of the gauginos and sfermions as a function
of the parameter $\delta$ in
Fig.{\ref{mass_spectrum_GUT_without_unif}}.  Here the other Yukawa
couplings have been set as $\lambda_Y=0.3$, $(\lambda_1,
\lambda_2,\lambda_3)=(0.2,0.1,0.16)$, $(f_1,f_2,f_3)=(0.5,0,0.6)$,
$k_{d1}=k_{l1}=0$ at the GUT scale and the strong SU(2) gauge coupling 
$g$ has been taken as $g=2 \pi$ at the messenger scale. The dynamical scale 
$\Lambda$ of
the strong SU(2) gauge interaction and the vacuum expectation values
$\langle Z_2 \rangle$ and $\langle Z_3 \rangle$ 
have been taken as $\Lambda=1.0
\times 10^6~\GEV$ and $\langle Z_2 \rangle=\langle Z_3 \rangle=5.0
\times10^5~\GEV$, respectively. We have taken the parameters 
$k_2=0$, $k_3=0.8$, and $f=\bar{f}=0.2$
in Eq.(\ref{initial_nonunif}).  
\begin{figure}
  \centerline{\psfig{file=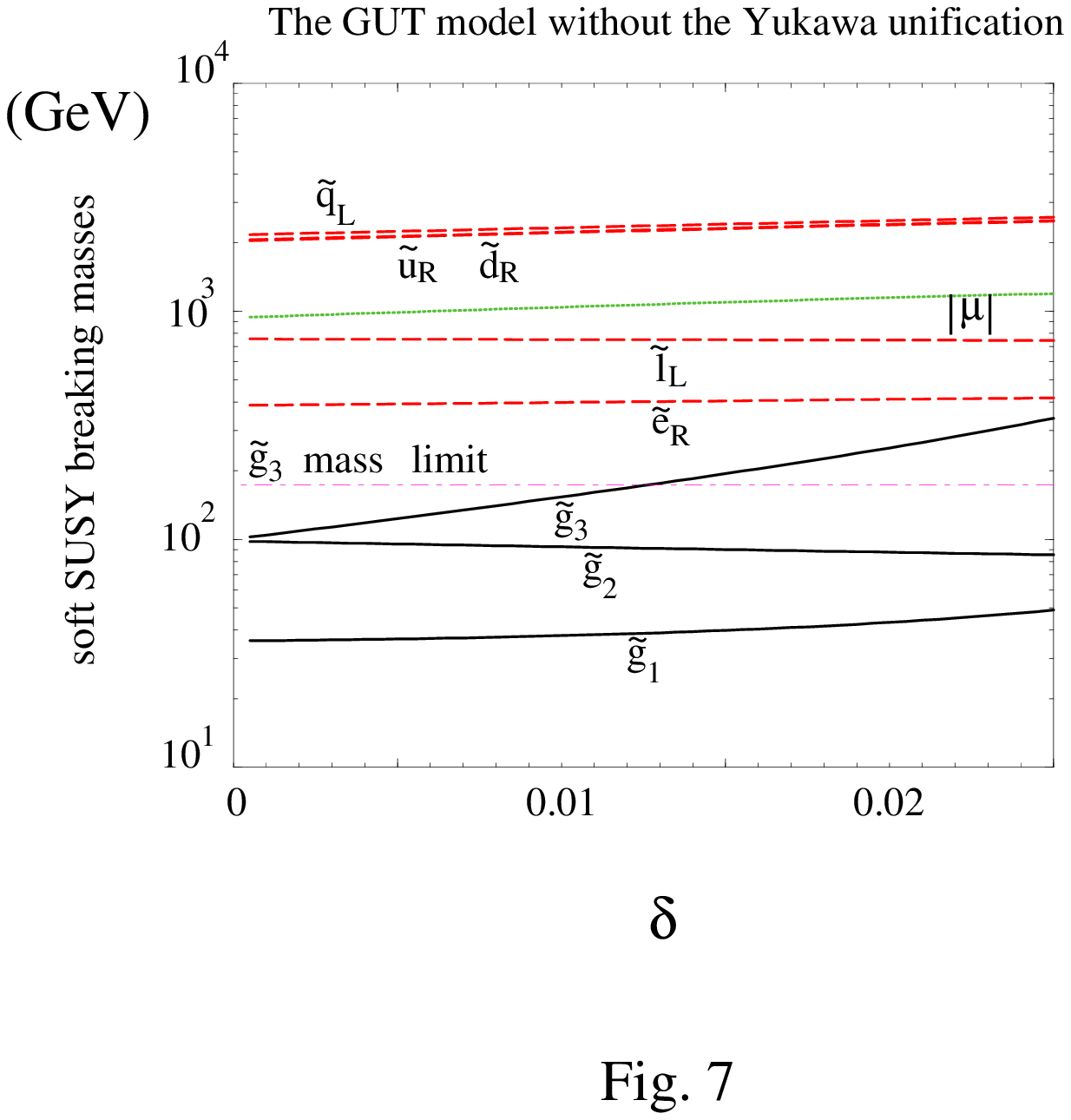,width=15cm}}
\caption{The mass spectrum of the gauginos and sfermions
  as a function of the parameter $\delta$. Here we have taken the Yukawa
  couplings at the GUT scale as follows: $\lambda_Y=0.3$, $(\lambda_1,
  \lambda_2,\lambda_3)=(0.2,0.1,0.16)$, $(f_1,f_2,f_3)=(0.5,0,0.6)$,
  $k_{d1}=k_{l1}=0$. We have set $k_2=0$, $k_3=0.8$, and
  $f=\bar{f}=0.2$ in Eq.(\ref{initial_nonunif}).
  Solid lines represent the gaugino
  masses: the bino $\tilde{g}_1$, wino $\tilde{g}_2$, and
  gluino $\tilde{g}_3$ masses, and dashed lines represent the sfermion
  masses: the left-handed squark $\tilde{q}_L$, right-handed
  sup $\tilde{u}_R$, right-handed sdown $\tilde{d}_R$, doublet
  slepton $\tilde{l}_L$, and
  right-handed selectron $\tilde{e}_R$ masses.  
  The $\mu$-parameter is also shown (dotted line).
  The renormalization effects from the messenger scale 
  to the electroweak scale have been taken into account.
  We also show the experimental
  lower bound on the gluino mass (dash-dotted line).}
\label{mass_spectrum_GUT_without_unif}
\end{figure}
From Fig.{\ref{mass_spectrum_GUT_without_unif}}, we see that the case 
with the Yukawa unification (i.e. $\delta=0$) is excluded by the experimental
lower bound on the gluino mass. However, the small corrections 
($\delta \sim 0.013$) from the
non-renormalizable terms may significantly change the mass spectrum.
This is because the
parameter $F^{(d)}/m_d m_{\bar{d}}$ gets larger than that in the case
with the Yukawa unification while $F^{(l)}/m_l m_{\bar{l}}$ being kept
unchanged.  Since the suppression factor in the gluino mass gets
closer to $1$, the gluino becomes heavier to escape 
the experimental lower bound.

It is relatively difficult to predict the mass spectrum in
the presence of the non-renormalizable terms. However, we should stress
that this model still has many distinguishable features 
as discussed in the previous sections: the GUT relation among
the gaugino masses does not hold, and the gauginos tend to be lighter
than the sfermions since the gaugino masses have the suppression
factor.

\section{Other interesting features in the present model}

From the naturalness point of view as discussed in
Section~\ref{RAD_section}, much large sfermion masses compared with the weak
scale are not preferable.  Then, it is natural to have a relatively lower SUSY
breaking scale, that is, the lighter gravitino which satisfies the
cosmological requirement $m_{3/2} <1~\KEV$.  
Such a light gravitino may bring us other
interesting consequences in the collider experiments.  When the
gravitino is so light that the cosmological requirement is satisfied,
the gravitino becomes the LSP.  Thus the NLSP, primarily the bino,
decays into the gravitino emitting a photon. 
The goldstino component of the gravitino has the following
interaction with the matters \cite{goldstino_intaraction},
\begin{eqnarray}
  {\cal{L}} &=& - \frac{1}{F} j^{\alpha \mu} \partial_\mu
  {\tilde{G}_\alpha}, \nonumber \\ &=& \cos \theta
  \frac{m_{\tilde{g}_1}}{2 \sqrt{2} F} \tilde{B} \bar{\sigma}^\mu
  \sigma^\nu \tilde{G} F_{\mu \nu} + \cdots.
\label{goldstino_int}
\end{eqnarray}
An important point is that the couplings of the goldstino are
suppressed by only the SUSY breaking scale, not the Planck scale. Therefore,
the couplings of the goldstino get larger as the SUSY breaking scale
becomes smaller.  If the SUSY breaking scale is sufficiently small,
the decay of the superparticles into the gravitino may occur within
the detector.

From the interaction Eq.(\ref{goldstino_int}), we find the decay rate
of the bino as 
\begin{eqnarray}
  \Gamma(\tilde{g}_1 \rightarrow \tilde{G} \gamma )= \frac{
    m^5_{\tilde{g}_1} \cos^2 \theta_W}{16 \pi F^2}.
\label{decay_width_bino}
\end{eqnarray}
Then the decay length $L$ of the bino with energy $E$ in the laboratory frame
is given by
\begin{eqnarray}
  L = 1.3 \left( \frac{100~\GEV}{m_{\tilde{g}_1}} \right)^5
\left( \frac{\sqrt{F}}{10^6~\GEV} \right)^4
\left(\frac{E^2}{m_{\tilde{g}_1}^2}-1 \right)^{\frac{1}{2}}~~{\rm m}.
\label{decay_length}
\end{eqnarray}
For example, when $\sqrt{F}=5\times 10^5~\GEV$ which corresponds
to the gravitino mass with $m_{3/2}=0.06~\KEV$ and $m_{\tilde{g}_1}=100~\GEV$,
the decay length of the bino is $8 {\rm cm}$ for 
$(E^2/{m_{\tilde{g}_1}^2}-1)^{1/2}=1$.
Since it is possible for the gravitino mass to be much lighter than
$m_{3/2} <1~\KEV$, we may find signature
``$\gamma \gamma +$ missing energy'' in the collider experiments.
We should notice that any realistic models with a sizable  
signature ``$\gamma \gamma +$ missing energy'' 
have not been known except for the present model.
The phenomenological investigations have been done in many literatures
\cite{Rad_sym_break_MMM1, gg_missing} only under the assumption of 
the existence of very
light gravitino.\footnote{ The signature
  ``$\gamma \gamma +$ missing energy'' does not necessarily suggest
  the light gravitino. It has been known that the light axino in the
  framework of the no-scale supergravity also brings us such a signal
  in the collider experiments \cite{HTY}. Therefore, other experiments,
  axion search and direct superparticle search for example, are also
  necessary to distinguish the present model from other models.}

Recently, however, the constraints on the SUSY models with 
``$\gamma \gamma +$ missing energy'' signal have been reported
in Ref. \cite{ELN}. The lower bounds on the lightest neutralino
and the lightest chargino masses are obtained as
\begin{eqnarray}
m_{\tilde{g}_1} &\gsim& 75~\GEV, \nonumber \\  
m_{\tilde{g}_2} &\gsim& 150~\GEV,
\label{gg_bound}
\end{eqnarray}
where the GUT relation among the gaugino masses is assumed.
To satisfy the mass bounds Eq.(\ref{gg_bound}), 
the squarks and sleptons should be very heavy ($m_{\tilde{f}}\sim 10~\TEV$)
or the parameter $F^{(\psi)}/m_\psi m_{\bar{\psi}}$ should be very
close to $1$ in the present model.
If the squarks and sleptons are so heavy, 
$m_{\tilde{f}}\sim 10~\TEV$, a fine tuning may be needed to
obtain the correct electroweak scale as discussed in section
\ref{RAD_section}. On the other hand, in the case where 
the parameter $F^{(\psi)}/m_\psi m_{\bar{\psi}}$  is very
close to $1$, it already means a fine tuning.
From the naturalness point of view, we may exclude the parameter regions
where the ``$\gamma \gamma +$ missing energy'' signal is detectable 
in the present experiments.

The decay length Eq.(\ref{decay_length}) strongly depends on the bino
mass as well as the SUSY breaking scale.  For example, when $\sqrt{F}
=5\times10^5~\GEV$ and $m_{\tilde{g}_1}=40~\GEV$, $L=8$ m for 
$(E^2/{m_{\tilde{g}_1}^2}-1)^{1/2}=1$ which is
larger than the typical detector size.  Therefore, the signal
``$\gamma \gamma +$ missing energy'' can not be observed within the
detector when the bino mass is much smaller than $100~\GEV$.  In the
case where the bino slowly decays outside the detector, the bounds in
Eq.(\ref{gg_bound}) are not applicable because they are derived under
the assumption that the bino completely decays into the 
photon and gravitino within the detector.  
Since our model suggests the light gauginos as discussed in
the previous sections, it is likely the case.  Although the 
``$\gamma \gamma +$ missing energy'' signature is not expected in the
existing experiments, the pair production of  
the light gauginos can be observed directly in the future
collider experiments. Furthermore, if we require that 
$m_{3/2}~\lsim~1~\KEV$ or $m_{\tilde{f}}~\lsim~10~\TEV$, 
the SUSY breaking scale becomes as $\sqrt{F}~\lsim~2 \times 
10^{6}~\GEV$.
Then, according to Ref. \cite{Orito}, the slow decay of the bino 
may be detectable in the near future experiments 
such as LHC, even for the case of long $L$.\footnote{In the experiment
proposed in Ref. \cite{Orito}, the ``$\gamma \gamma +$ missing energy'' 
signature can be detectable at LHC as long as $\sqrt{F}~\lsim~10^7~\GEV$.
Thus such a signal will be observed in our model.
In other models which have the large $\sqrt{F} > 10^7~\GEV$, 
however, this signal can not be observable at LHC even with the same $L$.}

Finally we remark on the unification of the gauge coupling constants.
To break the SUSY dynamically, we assume the strong SU(2) gauge 
interaction with the four fundamental representation fields.
It is remarkable that all gauge coupling constants, not only
SU(3)$_C~\times$ SU(2)$_L~\times$ U(1)$_Y$ gauge couplings but also
strong SU(2) gauge coupling, meet at the scale $\sim 10^{16}~\GEV$
when we set the strong SU(2) dynamical scale $\Lambda$ as
$\Lambda\simeq 10^{5-6}~\GEV$ in order that 
$m_{3/2} \sim 0.1-1~\KEV$ \cite{INTY}.

\section{Conclusion}

We have performed a detailed analysis of 
a direct-transmission model of the dynamical SUSY
breaking previously proposed in Ref. \cite{INTY}.  This model 
possesses many remarkable points: there are no SUSY FCNC , 
all mass scales are generated from the strong SU(2) dynamics ($\mu$-term
may also originate from the same dynamics), and the unification of all
gauge coupling constants of SU(3)$_C$, SU(2)$_L$, U(1)$_Y$, and the
strong SU(2) gauge groups may be realized. Furthermore it is 
quite natural for the gravitino mass to be smaller than $1~\KEV$ as
required from the standard cosmology. We notice that 
this cosmological requirement is not satisfied by any other models 
which have been proposed.

In the present model, there are many distinguishable low-energy features
from in other models: the so-called GUT relation among the gaugino masses
does not hold even if we consider the GUT models. Furthermore, the
gauginos become lighter than sfermions because the gaugino masses
have extra suppression factors.  When the suppression of the 
gaugino masses is large, squarks tend to be heavy in order to 
satisfy the experimental lower bounds on the gaugino masses.
In this case, the radiative electroweak symmetry breaking requires 
a large $\mu$-parameter, and we may need a fine tuning.
If we consider the GUT model with the
Yukawa unification, the gluino mass tends to be lighter than the wino
mass. However, the Yukawa unification is broken in the presence of
non-renormalizable interactions. Then the prediction of the mass
spectrum of the superparticles is weakened.

Moreover, the light gravitino brings us 
a fascinating signature, that is, ``$\gamma \gamma+$ missing energy'' 
in the collider experiments when the bino is relatively heavy 
($m_{\tilde{g}_1}~\gsim~100~\GEV$).  
From the naturalness point of view,
however, the light gauginos are most likely in our model. 
Thus the collider signature will not be observed in the present 
experiments because the bino decays outside the detector. 
Even in this case, 
the bino decay may be detectable \cite{Orito} 
in the near future colliders
because of the relatively lower SUSY breaking scale.

\section*{Acknowledgment}
We would like to thank Izawa K.-I. and T.~Yanagida for very useful
comments and careful reading of the manuscript.

\appendix
\section{Calculation of the gaugino and sfermion masses}

In this Appendix, we show the detailed calculation of the gaugino
and sfermion masses in the MSSM sector.

When we take the mass eigenbasis in the messenger fermions and sfermions as
discussed in section \ref{Mass_spectrum},
gaugino-fermion-scalar interaction is 
\begin{eqnarray}
  {\cal{L}}_{s}&=&\sqrt{2}ig\sum_{\alpha = 1}^{2}\sum_{\beta = 1}^{4}
  [(T_{\beta 1}^{(\psi)}V_{1 \alpha}^{(\psi) \dagger} + T_{\beta
    2}^{(\psi)}V_{2 \alpha}^{(\psi) \dagger})\tilde{\psi}_{i\beta}^{*}
  {\rm T^{(a)}_{ij}}\psi_{j\alpha}{\tilde{g}}^{(a)} \nonumber \\ & &
  \qquad -(V_{\alpha 1}^{(\psi)}T_{1 \beta}^{(\psi) \dagger} +
  V_{\alpha 2}^{(\psi)}T_{2 \beta}^{(\psi) \dagger}){\tilde{g}}^{(a)*}
  \psi_{i\alpha}^{*}{\rm T^{(a)}_{ij}}\tilde{\psi}_{j\beta} \nonumber
  \\ & & \qquad -(U_{\alpha 1}^{(\psi)}T_{3 \beta}^{(\psi) \dagger} +
  U_{\alpha 2}^{(\psi)}T_{4 \beta}^{(\psi)
    \dagger})\bar{\psi}_{i\alpha} {\rm
    T^{(a)}_{ij}}\tilde{\psi}_{j\beta}{\tilde{g}}^{(a)} \nonumber \\ &
  & \qquad +(T_{\beta 3}^{(\psi)}U_{1 \alpha}^{(\psi) \dagger} +
  T_{\beta 4}^{(\psi)}U_{2 \alpha}^{(\psi) \dagger}){\tilde{g}}^{(a)*}
\tilde{\psi}_{i\beta}^{*}{\rm T^{(a)}_{ij}}\bar{\psi}_{j\alpha}^{*}],
\end{eqnarray}
where $\psi_\alpha$ and $\bar{\psi}_\alpha$ denote the messenger fermions in 
the mass eigenstates and
$\tilde{\psi}_\alpha$ is the messenger sfermions in the mass eigenstates
as follows:
\begin{eqnarray}
\left(
\begin{array}{c}
  \psi_1 \\ 
\psi_2
\end{array}
\right)
=V^{(\psi)}
\left(
\begin{array}{c}
  \psi \\ 
\psi'
\end{array}
\right),
\\
\left(
\begin{array}{c}
  \bar{\psi}_1 \\ 
\bar{\psi}_2
\end{array}
\right)
=U^{(\psi)*}
\left(
\begin{array}{c}
  \bar{\psi} \\ 
\bar{\psi}'
\end{array}
\right),
\\
\left(
\begin{array}{c}
  \tilde{\psi}_1 \\ \tilde{\psi}_2 \\ \tilde{\psi}_3 \\ 
\tilde{\psi}_4 
\end{array}
\right)
=T^{(\psi)}
\left(
\begin{array}{c}
  \tilde{\psi} \\ \tilde{\psi}' \\ \tilde{\bar{\psi}}^{*} \\ 
\tilde{\bar{\psi}}'^{*} \\
\end{array}
\right),
\end{eqnarray}
and 
${\tilde{g}}^{(a)}$ is the gaugino and ${\rm T^{(a)}_{ij}}$ represents
the generator of the gauge group. Then the MSSM gaugino masses
coming from Fig.\ref{gaugino_mass_graph}
becomes 
\begin{eqnarray}
  m_{\tilde{g}_3} &=& \frac{\alpha_3}{2\pi} {\cal F}^{(d)},
\\
m_{\tilde{g}_2} &=& \frac{\alpha_2}{2\pi} {\cal F}^{(l)},
\\
m_{\tilde{g}_1} &=& \frac{\alpha_1}{2\pi} \left\{
\frac{2}{5} {\cal F}^{(d)} + \frac{3}{5} {\cal F}^{(l)} \right\},
\end{eqnarray}
where the masses $m_{\tilde{g}_i}$ ($i=1, \cdots, 3$) denote the bino,
wino, and gluino masses, respectively, and we have adopted the SU(5)
GUT normalization of the U(1)$_Y$ gauge coupling ($\alpha_1 \equiv
\frac{5}{3} \alpha_Y$).
The functions ${\cal F}^{(\psi)}$ are given by 
\begin{eqnarray}
  {\cal F}^{(\psi)} &=& 32\pi^{2}i\sum_{\alpha = 1}^{2}\sum_{\beta =
    1}^{4} (U_{\alpha 1}^{(\psi)}T_{3 \beta}^{(\psi) \dagger} +
  U_{\alpha 2}^{(\psi)}T_{4 \beta}^{(\psi) \dagger}) (T_{\beta
    1}^{(\psi)}V_{1 \alpha}^{(\psi) \dagger} + T_{\beta
    2}^{(\psi)}V_{2 \alpha}^{(\psi) \dagger}) \nonumber \\ & & \qquad
  \times {\rm T^{(a)}_{ij}}{\rm T^{(a)}_{ji}}M^{(\psi)}_{\alpha}
  \int\frac{d^{4}k}{(2\pi)^4}\frac{1}{(p-k)^2+m^{(\psi)2}_{\beta}}
  \frac{1}{k^2+M^{(\psi)2}_{\alpha}}|_{p=0} \nonumber \\ &=&
  16\pi^{2}i\sum_{\alpha = 1}^{2}\sum_{\beta = 1}^{4} (U_{\alpha
    1}^{(\psi)}T_{3 \beta}^{(\psi) \dagger} + U_{\alpha
    2}^{(\psi)}T_{4 \beta}^{(\psi) \dagger}) (T_{\beta 1}^{(\psi)}V_{1
    \alpha}^{(\psi) \dagger} + T_{\beta 2}^{(\psi)}V_{2
    \alpha}^{(\psi) \dagger}) \nonumber \\ & & \qquad \times
  M^{(\psi)}_{\alpha} \{ -\frac{i}{16\pi^{2}}\int_{0}^{1}dx\ln
  [xm^{(\psi)2}_{\beta} +(1-x)M^{(\psi)2}_{\alpha}-p^2 x(1-x)]
  \}|_{p=0} \nonumber \\ &=& \sum_{\alpha =
    1}^{2}\sum_{\beta = 1}^{4} M^{(\psi)}_{\alpha}(U_{\alpha 1}^{(\psi)}T_{3
    \beta}^{(\psi) \dagger} + U_{\alpha 2}^{(\psi)}T_{4 \beta}^{(\psi)
    \dagger}) (T_{\beta 1}^{(\psi)}V_{1 \alpha}^{(\psi) \dagger} +
  T_{\beta 2}^{(\psi)}V_{2 \alpha}^{(\psi) \dagger}) \nonumber \\ & &
  \qquad \times \int_{0}^{1}dx\ln
  [xm^{(\psi)2}_{\beta}+(1-x)M^{(\psi)2}_{\alpha}] \nonumber \\ &=&
  \sum_{\alpha = 1}^{2}\sum_{\beta = 1}^{4}M^{(\psi)}_{\alpha}
  (U_{\alpha 1}^{(\psi)}T_{3 \beta}^{(\psi) \dagger} + U_{\alpha
    2}^{(\psi)}T_{4 \beta}^{(\psi) \dagger}) (T_{\beta 1}^{(\psi)}V_{1
    \alpha}^{(\psi) \dagger} + T_{\beta 2}^{(\psi)}V_{2
    \alpha}^{(\psi) \dagger}) \nonumber \\ & & \qquad \times
  \frac{m^{(\psi)2}_{\beta}}
  {m^{(\psi)2}_{\beta}-M^{(\psi)2}_{\alpha}}
\ln\frac{m^{(\psi)2}_{\beta}}{M^{(\psi)2}_{\alpha}}. 
\end{eqnarray}
We next consider the squark and slepton masses.
These masses arise from Fig.\ref{sfermion_mass_graph} and are given by 
\begin{eqnarray}
  m^2_{\tilde{f}}=\frac{1}{2} \left[ C_3^{\tilde{f}}
  \left(\frac{\alpha_3}{4 \pi} \right)^2 {\cal G}^{(d)2} +
    C_2^{\tilde{f}} \left(\frac{\alpha_2}{4 \pi} \right)^2 {\cal
      G}^{(l)2} + \frac{3}{5} Y^2 \left(\frac{\alpha_1}{4 \pi}
  \right)^2 \left(\frac{2}{5} {\cal G}^{(d)2}
  + \frac{3}{5} {\cal G}^{(l)2} \right) \right],
\end{eqnarray}
where $C_3^{\tilde{f}}=\frac{4}{3}$ and $C_2^{\tilde{f}}=\frac{3}{4}$
when $\tilde{f}$ is in the fundamental representation of SU(3)$_C$ and
SU(2)$_L$, and $C_i^{\tilde{f}}=0$ for the gauge singlets, and $Y$
denotes the U(1)$_Y$ hypercharge
($Y \equiv Q-T_3$). Here ${\cal G}^{(\psi)2}$ are given by \cite{Martin} 
\begin{eqnarray}
  {\cal G}^{(\psi)2} &=& (4\pi)^4 \{ -\sum_{\alpha =
    1}^{4}<m_{\alpha}^{(\psi)}|m_{\alpha}^{(\psi)}|0> -4\sum_{\alpha =
    1}^{4}<m_{\alpha}^{(\psi)}|m_{\alpha}^{(\psi)}|0,0> \nonumber \\ &
  & \qquad -4\sum_{\alpha =
    1}^{2}<M_{\alpha}^{(\psi)}|M_{\alpha}^{(\psi)}|0> +8\sum_{\alpha =
    1}^{2}<M_{\alpha}^{(\psi)}|M_{\alpha}^{(\psi)}|0,0> \nonumber \\ &
  & \qquad -\sum_{\alpha = 1}^{4} \sum_{\beta = 1}^{4} (T_{\alpha
    1}^{(\psi)}T_{1 \beta}^{(\psi) \dagger} + T_{\alpha
    2}^{(\psi)}T_{2 \beta}^{(\psi) \dagger} - T_{\alpha
    3}^{(\psi)}T_{3 \beta}^{(\psi) \dagger} - T_{\alpha
    4}^{(\psi)}T_{4 \beta}^{(\psi) \dagger}) \nonumber \\ & & \qquad
  \qquad \times (T_{\beta 1}^{(\psi)}T_{1 \alpha}^{(\psi) \dagger} +
  T_{\beta 2}^{(\psi)}T_{2 \alpha}^{(\psi) \dagger} - T_{\beta
    3}^{(\psi)}T_{3 \alpha}^{(\psi) \dagger} - T_{\beta
    4}^{(\psi)}T_{4 \alpha}^{(\psi) \dagger})
  <m_{\alpha}^{(\psi)}|m_{\beta}^{(\psi)}|0> \nonumber \\ & & \qquad
  +4\sum_{\alpha = 1}^{4} \sum_{\beta = 1}^{2} [(V_{\beta
    1}^{(\psi)}T_{1 \alpha}^{(\psi) \dagger} + V_{\beta
    2}^{(\psi)}T_{2 \alpha}^{(\psi) \dagger}) (T_{\alpha
    1}^{(\psi)}V_{1 \beta}^{(\psi) \dagger} + T_{\alpha
    2}^{(\psi)}V_{2 \beta}^{(\psi) \dagger}) \nonumber \\ & & \qquad
  \qquad + (U_{\beta 1}^{(\psi)}T_{3 \alpha}^{(\psi) \dagger} +
  U_{\beta 2}^{(\psi)}T_{4 \alpha}^{(\psi) \dagger}) (T_{\alpha
    3}^{(\psi)}U_{1 \beta}^{(\psi) \dagger} + T_{\alpha
    4}^{(\psi)}U_{2 \beta}^{(\psi) \dagger})] \nonumber \\ & & \qquad
  \qquad [<m_{\alpha}^{(\psi)}|M_{\beta}^{(\psi)}|0>
  +(m_{\alpha}^{(\psi)2}-M_{\beta}^{(\psi)2})
<m_{\alpha}^{(\psi)}|M_{\beta}^{(\psi)}|0,0>] \} ,
\end{eqnarray}
where
\begin{eqnarray}
  <m_{1}|m_{2}|m_{3}>&=&\int\frac{d^4 q}{(2\pi)^4}\frac{d^4
    k}{(2\pi)^4} \frac{1}{(q^2 +m_1^2 )(k^2 +m_2^2 )([k-q]^2 +m_3^2 )} ,
  \\ <m_{1}|m_{2}|m_{3},m_{3}>&=&\int\frac{d^4 q}{(2\pi)^4}\frac{d^4
    k}{(2\pi)^4}
\frac{1}{(q^2 +m_1^2 )(k^2 +m_2^2 )([k-q]^2 +m_3^2 )^2} .
\end{eqnarray}
This becomes 
\begin{eqnarray}
  {\cal G}^{(\psi)2} &=& \sum_{\alpha = 1}^{4} m_{\alpha}^{(\psi)2} (4
  \ln m_{\alpha}^{(\psi)2} + \ln^{2}m_{\alpha}^{(\psi)2}) +
  \sum_{\alpha = 1}^{2} M_{\alpha}^{(\psi)2} (-8 \ln
  M_{\alpha}^{(\psi)2} + 4 \ln^{2}M_{\alpha}^{(\psi)2}) \nonumber \\ &
  & + \sum_{\alpha = 1}^{4} \sum_{\beta = 1}^{4} (T_{\alpha
    1}^{(\psi)}T_{1 \beta}^{(\psi) \dagger} + T_{\alpha
    2}^{(\psi)}T_{2 \beta}^{(\psi) \dagger} - T_{\alpha
    3}^{(\psi)}T_{3 \beta}^{(\psi) \dagger} - T_{\alpha
    4}^{(\psi)}T_{4 \beta}^{(\psi) \dagger}) \nonumber \\ & & \qquad
  \times (T_{\beta 1}^{(\psi)}T_{1 \alpha}^{(\psi) \dagger} + T_{\beta
    2}^{(\psi)}T_{2 \alpha}^{(\psi) \dagger} - T_{\beta
    3}^{(\psi)}T_{3 \alpha}^{(\psi) \dagger} - T_{\beta
    4}^{(\psi)}T_{4 \alpha}^{(\psi) \dagger}) \nonumber \\ & & \qquad
  \times m_{\alpha}^{(\psi)2} [-\ln^{2}m_{\beta}^{(\psi)2} + 2\ln
  m_{\alpha}^{(\psi)2}\ln m_{\beta}^{(\psi)2} - 2{\rm Li}_2
  (1-\frac{m_{\alpha}^{(\psi)2}}{m_{\beta}^{(\psi)2}})] \nonumber \\ &
  & + 2 \sum_{\alpha = 1}^{4} \sum_{\beta = 1}^{2} [(V_{\beta
    1}^{(\psi)}T_{1 \alpha}^{(\psi) \dagger} + V_{\beta
    2}^{(\psi)}T_{2 \alpha}^{(\psi) \dagger}) (T_{\alpha
    1}^{(\psi)}V_{1 \beta}^{(\psi) \dagger} + T_{\alpha
    2}^{(\psi)}V_{2 \beta}^{(\psi) \dagger}) \nonumber \\ & & \qquad
  \qquad + (U_{\beta 1}^{(\psi)}T_{3 \alpha}^{(\psi) \dagger} +
  U_{\beta 2}^{(\psi)}T_{4 \alpha}^{(\psi) \dagger}) (T_{\alpha
    3}^{(\psi)}U_{1 \beta}^{(\psi) \dagger} + T_{\alpha
    4}^{(\psi)}U_{2 \beta}^{(\psi) \dagger})] \nonumber \\ & & \qquad
  \times [m_{\alpha}^{(\psi)2} (\ln^{2}M_{\beta}^{(\psi)2} -\ln
  m_{\alpha}^{(\psi)2}\ln M_{\beta}^{(\psi)2} +{\rm Li}_2
  (1-\frac{m_{\alpha}^{(\psi)2}}{M_{\beta}^{(\psi)2}}) -{\rm Li}_2
  (1-\frac{M_{\beta}^{(\psi)2}}{m_{\alpha}^{(\psi)2}})) \nonumber \\ &
  & \qquad + M_{\beta}^{(\psi)2} (\ln^{2}m_{\alpha}^{(\psi)2} -\ln
  m_{\alpha}^{(\psi)2}\ln M_{\beta}^{(\psi)2} +{\rm Li}_2
  (1-\frac{M_{\beta}^{(\psi)2}}{m_{\alpha}^{(\psi)2}}) +{\rm Li}_2
  (1-\frac{m_{\alpha}^{(\psi)2}}{M_{\beta}^{(\psi)2}}))].
\nonumber \\
\end{eqnarray}
Here ${\rm Li}_2 (x) =
-\int_{0}^{1}(dt/t)\ln (1-xt)$ is a dilog function.

\section{Renormalization group equations for the model}

In this Appendix, we list the RGEs for the model.

The superpotential at the scale above $\Lambda$ is 
\begin{eqnarray}
  W &=& \frac{\lambda}{\sqrt{2}}\epsilon_{\alpha\beta}\{
  Q^{\alpha}_{1}Q^{\beta}_{3}Y_{1}+Q^{\alpha}_{1}Q^{\beta}_{4}Y_{2}+
  Q^{\alpha}_{2}Q^{\beta}_{3}Y_{3}+Q^{\alpha}_{2}Q^{\beta}_{4}Y_{4}+
  \frac{1}{\sqrt{2}}(Q^{\alpha}_{1}Q^{\beta}_{2}-Q^{\alpha}_{3}Q^{\beta}_{4})
  Y_{5}\} \nonumber \\ & & \quad
  +\frac{\lambda_{1}}{\sqrt{2}}\epsilon_{\alpha\beta}
  (Q^{\alpha}_{1}Q^{\beta}_{2}+Q^{\alpha}_{3}Q^{\beta}_{4})Z_{1}+
  k_{d1}d\bar{d}Z_{1}+k_{l1}l\bar{l}Z_{1} -f_{1}X^{2}Z_{1} \nonumber
  \\ & & \quad +\frac{\lambda_{2}}{\sqrt{2}}\epsilon_{\alpha\beta}
  (Q^{\alpha}_{1}Q^{\beta}_{2}+Q^{\alpha}_{3}Q^{\beta}_{4})Z_{2}+
  k_{d2}d\bar{d}Z_{2}+k_{l2}l\bar{l}Z_{2} -f_{2}X^{2}Z_{2} \nonumber
  \\ & & \quad +\frac{\lambda_{3}}{\sqrt{2}}\epsilon_{\alpha\beta}
  (Q^{\alpha}_{1}Q^{\beta}_{2}+Q^{\alpha}_{3}Q^{\beta}_{4})Z_{3}+
  k_{d3}d\bar{d}Z_{3}+k_{l3}l\bar{l}Z_{3} -f_{3}X^{2}Z_{3} \nonumber
  \\ & & \quad +f_{d}d{\bar d}'X+f_{\bar d}d'{\bar d}X
+f_{l}l{\bar l}'X+f_{\bar l}l'{\bar l}X.
\end{eqnarray}
Then, one-loop RGEs for the Yukawa
coupling constants are given by 
\begin{eqnarray*}
  \mu\frac{d\lambda}{d\mu} &=& \frac{1}{16\pi^{2}}
  \frac{1}{2}\lambda(\lambda_{1}^{2}+\lambda_{2}^{2}+\lambda_{3}^{2}
  +7\lambda^{2}-6g^{2}), \nonumber \\ \nonumber \\ 
  \mu\frac{d\lambda_{1}}{d\mu} &=& \frac{1}{16\pi^{2}}
  [\frac{1}{2}\lambda_{1}(\lambda_{1}^{2}+\lambda_{2}^{2}+\lambda_{3}^{2}
  +5\lambda^{2}-6g^{2}) \nonumber \\ & & \qquad \qquad \qquad \qquad
  \qquad +\lambda_{i}(\lambda_{i}\lambda_{1}
  +3k_{di}k_{d1}+2k_{li}k_{l1}+2f_{i}f_{1})], \nonumber \\ 
  \mu\frac{dk_{d1}}{d\mu} &=& \frac{1}{16\pi^{2}}
  [k_{d1}(2k_{d1}^{2}+2k_{d2}^{2}+2k_{d3}^{2}+f_{d}^{2}+f_{\bar d}^{2}
  -\frac{16}{3}g_{3}^{2}-\frac{4}{15}g_{1}^{2}) \nonumber \\ & &
  \qquad \qquad \qquad \qquad \qquad
  +k_{di}(\lambda_{i}\lambda_{1}+3k_{di}k_{d1}+2k_{li}k_{l1}
  +2f_{i}f_{1})], \nonumber \\ \mu\frac{dk_{l1}}{d\mu} &=&
  \frac{1}{16\pi^{2}}
  [k_{l1}(2k_{l1}^{2}+2k_{l2}^{2}+2k_{l3}^{2}+f_{l}^{2}+f_{\bar l}^{2}
  -3g_{2}^{2}-\frac{3}{5}g_{1}^{2}) \nonumber \\ & & \qquad \qquad
  \qquad \qquad \qquad
  +k_{li}(\lambda_{i}\lambda_{1}+3k_{di}k_{d1}+2k_{li}k_{l1}
  +2f_{i}f_{1})], \nonumber \\ \mu\frac{df_{1}}{d\mu} &=&
  \frac{1}{16\pi^{2}} [2f_{1}(4f_{1}^{2}+4f_{2}^{2}+4f_{3}^{2}+
  3f_{d}^{2}+3f_{\bar d}^{2}+2f_{l}^{2}+2f_{\bar l}^{2}) \nonumber \\ 
  & & \qquad \qquad \qquad \qquad \qquad
  +f_{i}(\lambda_{i}\lambda_{1}+3k_{di}k_{d1}+2k_{li}k_{l1}
  +2f_{i}f_{1})], \nonumber \\ \nonumber \\ 
  \mu\frac{d\lambda_{2}}{d\mu} &=& \frac{1}{16\pi^{2}}
  [\frac{1}{2}\lambda_{2}(\lambda_{1}^{2}+\lambda_{2}^{2}+\lambda_{3}^{2}
  +5\lambda^{2}-6g^{2}) \nonumber \\ & & \qquad \qquad \qquad \qquad
  \qquad +\lambda_{i}(\lambda_{i}\lambda_{2}
  +3k_{di}k_{d2}+2k_{li}k_{l2}+2f_{i}f_{2})], \nonumber \\ 
  \mu\frac{dk_{d2}}{d\mu} &=& \frac{1}{16\pi^{2}}
  [k_{d2}(2k_{d1}^{2}+2k_{d2}^{2}+2k_{d3}^{2}+f_{d}^{2}+f_{\bar d}^{2}
  -\frac{16}{3}g_{3}^{2}-\frac{4}{15}g_{1}^{2}) \nonumber \\ & &
  \qquad \qquad \qquad \qquad \qquad
  +k_{di}(\lambda_{i}\lambda_{2}+3k_{di}k_{d2}+2k_{li}k_{l2}
  +2f_{i}f_{2})], \nonumber \\ \mu\frac{dk_{l2}}{d\mu} &=&
  \frac{1}{16\pi^{2}}
  [k_{l2}(2k_{l1}^{2}+2k_{l2}^{2}+2k_{l3}^{2}+f_{l}^{2}+f_{\bar l}^{2}
  -3g_{2}^{2}-\frac{3}{5}g_{1}^{2}) \nonumber \\ & & \qquad \qquad
  \qquad \qquad \qquad
  +k_{li}(\lambda_{i}\lambda_{2}+3k_{di}k_{d2}+2k_{li}k_{l2}
  +2f_{i}f_{2})], \nonumber \\ \mu\frac{df_{2}}{d\mu} &=&
  \frac{1}{16\pi^{2}} [2f_{2}(4f_{1}^{2}+4f_{2}^{2}+4f_{3}^{2}+
  3f_{d}^{2}+3f_{\bar d}^{2}+2f_{l}^{2}+2f_{\bar l}^{2}) \nonumber \\ 
  & & \qquad \qquad \qquad \qquad \qquad
  +f_{i}(\lambda_{i}\lambda_{2}+3k_{di}k_{d2}+2k_{li}k_{l2}
  +2f_{i}f_{2})], \nonumber \\ \nonumber \\ 
  \mu\frac{d\lambda_{3}}{d\mu} &=& \frac{1}{16\pi^{2}}
  [\frac{1}{2}\lambda_{3}(\lambda_{1}^{2}+\lambda_{2}^{2}+\lambda_{3}^{2}
  +5\lambda^{2}-6g^{2}) \nonumber \\ & & \qquad \qquad \qquad \qquad
  \qquad +\lambda_{i}(\lambda_{i}\lambda_{3}
  +3k_{di}k_{d3}+2k_{li}k_{l3}+2f_{i}f_{3})], \nonumber \\ 
  \mu\frac{dk_{d3}}{d\mu} &=& \frac{1}{16\pi^{2}}
  [k_{d3}(2k_{d1}^{2}+2k_{d2}^{2}+2k_{d3}^{2}+f_{d}^{2}+f_{\bar d}^{2}
  -\frac{16}{3}g_{3}^{2}-\frac{4}{15}g_{1}^{2}) \nonumber \\ & &
  \qquad \qquad \qquad \qquad \qquad
  +k_{di}(\lambda_{i}\lambda_{3}+3k_{di}k_{d3}+2k_{li}k_{l3}
  +2f_{i}f_{3})], \nonumber \\ \mu\frac{dk_{l3}}{d\mu} &=&
  \frac{1}{16\pi^{2}}
  [k_{l3}(2k_{l1}^{2}+2k_{l2}^{2}+2k_{l3}^{2}+f_{l}^{2}+f_{\bar l}^{2}
  -3g_{2}^{2}-\frac{3}{5}g_{1}^{2}) \nonumber \\ & & \qquad \qquad
  \qquad \qquad \qquad
  +k_{li}(\lambda_{i}\lambda_{3}+3k_{di}k_{d3}+2k_{li}k_{l3}
  +2f_{i}f_{3})], \nonumber \\ 
  \mu\frac{df_{3}}{d\mu}&=&\frac{1}{16\pi^{2}}
  [2f_{3}(4f_{1}^{2}+4f_{2}^{2}+4f_{3}^{2}+ 3f_{d}^{2}+3f_{\bar
    d}^{2}+2f_{l}^{2}+2f_{\bar l}^{2}) \nonumber \\ & & \qquad \qquad
  \qquad \qquad \qquad
  +f_{i}(\lambda_{i}\lambda_{3}+3k_{di}k_{d3}+2k_{li}k_{l3}
+2f_{i}f_{3})], \nonumber \\
%\nonumber \\
%
\end{eqnarray*}
\begin{eqnarray}
  \mu\frac{df_{d}}{d\mu} &=& \frac{1}{16\pi^{2}}f_{d}
  (k_{d1}^{2}+k_{d2}^{2}+k_{d3}^{2}+4f_{1}^{2}+4f_{2}^{2}+4f_{3}^{2}
  +5f_{d}^{2}+3f_{\bar d}^{2}+2f_{l}^{2}+2f_{\bar l}^{2}
  -\frac{16}{3}g_{3}^{2}-\frac{4}{15}g_{1}^{2}), \nonumber \\ 
  \mu\frac{df_{\bar d}}{d\mu} &=& \frac{1}{16\pi^{2}}f_{\bar d}
  (k_{d1}^{2}+k_{d2}^{2}+k_{d3}^{2}+4f_{1}^{2}+4f_{2}^{2}+4f_{3}^{2}
  +3f_{d}^{2}+5f_{\bar d}^{2}+2f_{l}^{2}+2f_{\bar l}^{2}
  -\frac{16}{3}g_{3}^{2}-\frac{4}{15}g_{1}^{2}), \nonumber \\ 
  \mu\frac{df_{l}}{d\mu} &=& \frac{1}{16\pi^{2}}f_{l}
  (k_{l1}^{2}+k_{l2}^{2}+k_{l3}^{2}+4f_{1}^{2}+4f_{2}^{2}+4f_{3}^{2}
  +3f_{d}^{2}+3f_{\bar d}^{2}+4f_{l}^{2}+2f_{\bar l}^{2}
  -3g_{2}^{2}-\frac{3}{5}g_{1}^{2}), \nonumber \\ \mu\frac{df_{\bar
      l}}{d\mu} &=& \frac{1}{16\pi^{2}}f_{\bar l}
  (k_{l1}^{2}+k_{l2}^{2}+k_{l3}^{2}+4f_{1}^{2}+4f_{2}^{2}+4f_{3}^{2}
  +3f_{d}^{2}+3f_{\bar d}^{2}+2f_{l}^{2}+4f_{\bar l}^{2}
-3g_{2}^{2}-\frac{3}{5}g_{1}^{2}), \nonumber \\ 
\end{eqnarray}
where $g , g_{3} , g_{2} , g_{1}$ denote strong SU(2), SU(3)$_{C}$ ,
SU(2)$_{L}$, U(1)$_{Y}$ gauge coupling constants respectively and 
the index $i$ is summed over $1,2,3$.

Further, one-loop RGEs for the gauge
coupling constants are given by 
\begin{eqnarray}
  \mu\frac{dg}{d\mu} &=& -\frac{4}{16\pi^{2}}g^{3}, \nonumber \\ 
  \mu\frac{dg_{3}}{d\mu} &=& -\frac{1}{16\pi^{2}}g_{3}^{3}, \nonumber
  \\ \mu\frac{dg_{2}}{d\mu} &=& \frac{3}{16\pi^{2}}g_{2}^{3}, \nonumber
  \\ 
\mu\frac{dg_{1}}{d\mu} &=& \frac{1}{16\pi^{2}}\frac{43}{5}g_{1}^{3}. 
\end{eqnarray}
\newpage
%
%%%%%%%%%%%%%%%%%%%%%%%%%%%%%%%%%%%%%%%%%%%%%%%%%%%%%%%%%%%%%%%
%
% NEW COMMANDS FOR THE BIBLIOGRAPHY
%
%%%%%%%%%%%%%%%%%%%%%%%%%%%%%%%%%%%%%%%%%%%%%%%%%%%%%%%%%%%%%%%
\newcommand{\Journal}[4]{{\sl #1} {\bf #2} {(#3)} {#4}}
\newcommand{\APJ}{Ap. J.} \newcommand{\CJP}{Can. J. Phys.}
\newcommand{\NC}{Nuovo Cimento} \newcommand{\NP}{Nucl. Phys.}
\newcommand{\PL}{Phys. Lett.} \newcommand{\PR}{Phys. Rev.}
\newcommand{\PRep}{Phys. Rep.} \newcommand{\PRL}{Phys. Rev. Lett.}
\newcommand{\PTP}{Prog. Theor. Phys.} \newcommand{\SJNP}{Sov. J. Nucl.
  Phys.}
\newcommand{\ZP}{Z. Phys.}
%%%%%%%%%%%%%%%%%%%%%%%%%%%%%%%%%%%%%%%%%%%%%%%%%%%%%%%%%%%%%%%

%
%
\end{document}